\documentclass[twocolumn]{aastex631}
\usepackage{color}
\usepackage{makecell}
\usepackage{appendix}
\usepackage{mathrsfs}

\usepackage{amsmath}

\shorttitle{Magnetization in Quasi-thermal Emission}
\shortauthors{Xin-Ying Song et al.}
\graphicspath{{./}{figures/}}
\begin{document}

\title{Spectral and Jet Properties of the Quasi-thermal Dominated GRB 210121A, GRB 210610B and GRB 221022B}

\correspondingauthor{Xin-Ying Song}%, Shuang-Nan Zhang}
\email{songxy@ihep.ac.cn}%}
%%
%% While authors can be grouped inside the same \author and \affiliation
%% commands it is better to have a single author for each. This allows for
%% one to exploit all the new benefits and should make book-keeping easier.
%%
%% If done correctly the peer review system will be able to
%% automatically put the author and affiliation information from the manuscript
%% and save the corresponding author the trouble of entering it by hand.

%\correspondingauthor{August Muench}
%\email{greg.schwarz@aas.org, gus.muench@aas.org}

\author[0000-0002-2176-8778]{Xin-Ying Song}
\affiliation{Key Laboratory of Particle Astrophysics, Institute of High Energy Physics, Chinese Academy of Sciences, Beijing 100049, China}
\author{Ling-Jun Wang}
\affil{Key Laboratory of Particle Astrophysics, Institute of High Energy Physics, Chinese Academy of Sciences, Beijing 100049, China}
\affil{University of Chinese Academy of Sciences, Chinese Academy of Sciences, Beijing 100049, China}
\author{Shu Zhang}
\affil{Key Laboratory of Particle Astrophysics, Institute of High Energy Physics, Chinese Academy of Sciences, Beijing 100049, China}
\affil{University of Chinese Academy of Sciences, Chinese Academy of Sciences, Beijing 100049, China}
\begin{abstract}
Some quasi-thermal (QT) dominated gamma-ray bursts (GRBs) could be well described by a multi-color blackbody (BB) function or a combined model of BB plus non-thermal (NT) component. In this analysis, two QT radiation-dominated bursts with known emission properties (GRB 210610B likely from a hybrid jet, and GRB 210121A with a spectrum consistent with a non-dissipative photospheric emission from a pure hot fireball) are used to make a comparison between these two modelings. To diagnose the magnetization properties of the central engine, the `top-down' approach proposed by Gao \& Zhang is adopted. It is found that diagnoses based on these two modelings could provide similar conclusions qualitatively; however, the modeling with mBB (or mBB+NT) may give more reasonable physical explanations. This implies that impacts from the GRB jet structure and the geometrical broadening on the observed spectrum should be considered. However, conservatively, these methods may be not sensitive enough to distinguish between the pure hot fireball and a mildly magnetized hybrid jet. Some other information is necessary to provide more evidence for the determination of jet properties for similar GRBs. Based on these considerations, we suggest that the photospheric emission of GRB 221022B is from a hot jet; a dissipation is caused by a internal shock (IS) mechanism due to the increasing Lorentz Factor with time, which makes its prompt emission behaves a typical evolution from thermal to NT.
\end{abstract}
%TC:endignore



\section{Introduction} \label{sec:intro}
The emission mechanism of gamma-ray bursts (GRBs) remains elusive despite about half a century of observation and investigation. There are two leading scenarios which have been suggested to interpret the observed spectra of GRBs. One is synchrotron radiation, which invokes a non-thermal emission of relativistic charged particles either from internal shocks (IS) or from internal magnetic dissipation processes \citep[e.g.,][]{2000ApJ...543..722L,2011ApJ...726...90Z}. The other is photospheric emission as a natural consequence of the fireball~\cite[e.g.,][]{1978MNRAS.183..359C,1986Are,1986ApJ308L43P}. However, the photospheric emission is not in its usual form because its spectrum is broadened. The Planck spectrum related to the photospheric emission could be broadened in two ways. First, dissipation below the photosphere can heat electrons above the equilibrium temperature. ISs below the photosphere~\citep[e.g.,][]{2005Dissipative}, magnetic reconnection~\citep[e.g.,][]{2004Spectra}, and hadronic collision shocks~\citep[e.g.,][]{2010Radiative} have been suggested as the dissipation processes. Second, the modification of Planck spectrum could be caused by geometrical broadening~\citep[e.g.][]{2008ApJ...682..463P}. This means that the observed spectrum is a superposition of a series of blackbodies of different temperatures, arising from different angles to the line of sight. Therefore, a multi-blackbody (mBB) function could be used to describe the quasi-thermal spectrum.

The possible origins for the observed photospheric emission in GRBs are listed as below: 
\begin{itemize}
    \item{pure hot fireball:} the pure hot fireball with or without dissipations; there is not Poynting flux  component in the outflow while the matter flux is dominant. From the above discussion, the spectrum of the quasi-thermal (QT) emission may be boarder than a blackbody (BB) function with a single temperature. GRB 210121A is a typical GRB from a pure hot fireball~\citep{2021ApJ...922..237W,2022ApJ...931..112S} and the spectrum is well consistent with the probability non-dissipative photosphere (NDP) model~\citep{2013MNRAS.428.2430L} from a pure hot fireball. In this case, the dimensionless entropy is $\eta=L_{\rm w}/\dot{M}c^2$ ($\dot{M}$ is the mass rate of baryon loading; $L_{\rm w}$ is the wind luminosity). %As shown in Figure~\ref{fig:spectra_magnetization} (a), the spectrum of mBB model are similar to that of the NDP model (their parameters are from the fit results to the spectrum of GRB 210121A in the first 0.9 s).They are both broader than that of the BB spectrum (with the temperature $kT= E_{\rm p}/\zeta$, $\zeta\sim3$ according to \cite{2012ApJ...758L..34Z}, and $E_{\rm p}$ is the peak energy of the $\nu F_{\nu}$ spectrum). This is mainly due to the geometrical broadening and the jet structure;
 \item{hybrid model:} a hybrid outflow in which the matter flux and Poynting flux both exist; in the case, we have $L_{\rm w}=L_{\rm m}+L_{\rm p}$ and
\begin{equation}\label{eq:L_sigma0}
    \eta(1+\sigma_0)=\frac{L_{\rm m}+L_{\rm p}}{\dot{M}c^2},
\end{equation}
 where the magnetization factor $\sigma_0$ is the ratio of Poynting flux luminosity to the matter flux; $L_{\rm p}$ is the luminosity of Poynting flux and $L_{\rm m}$ is the luminosity of  matter flux, and $\eta=\frac{L_{\rm m}}{\dot{M}c^2}$.  There are three different cases that may occur in a hybrid outflow~\citep{2015ApJ...801..103G}:
 \begin{itemize}
     \item{no sub-photosphere magnetic dissipation:} the Poynting flux component in the hybrid outflow could accelerate the outflow without any magnetic reconnection that occurs below the photosphere. The magnetic energy is only converted into the kinetic energy of the bulk motion. Such a scenario may be relevant to helical jets or self-sustained magnetic bubbles~\citep[e.g.][]{2001A&A...369..694S,2006ApJ...647.1192U,2012ApJ...757...56Y}. This scenario also predicts a QT photosphere emission component, which is consistent with the data of several Fermi GRBs~\citep[e.g.][]{2010ApJ...709L.172R,2012ApJ...757L..31A}; The spectrum from the outflow with a low to moderate magnetization ($\sigma_0<10$)~\citep{2022MNRAS.509.6047M}) with considering the  is numerically produced with considering the impacts from the GRB jet structure, the geometrical broadening and probability emission,%  as shown in Figure~\ref{fig:spectra_magnetization} (b),
     which is similar to the cases of pure hot fireball;
     \item{sub-photosphere magnetic dissipation:} the significant magnetic reconnection could occur below the photosphere; the photospheric emission is enhanced and could produce a spectrum with NT appearance and a larger $E_{\rm p}$ (1 MeV$\mbox{--}$20 MeV)~\citep{2013ApJ...764..157B,2015ApJ...802..134B}; 
     the Poynting flux could be thermalized completely below the photosphere due to the existence of the extra thermal component in the outflow. In this case, the spectrum may be similar to the case of pure hot fireball.
 \end{itemize}
\end{itemize}

 In the case of magnetic dissipation in the hybrid model, the calculation is quite complex if the complete thermalization below the photosphere is not achieved.  Since all of three GRBs analyzed in this work have lower $E_{\rm p}$ than that predicted in this case, we do not consider it in this work.  %Besides, if we consider an extremely case of II.C,   

\cite{2015ApJ...801..103G} proposes a `top-down' approach to diagnose the properties of the magnetization of the central engine, based on the observed quasi-thermal
photosphere emission properties. In this approach, these impacts from the jet structure and geometrical broadening on the spectrum shape are not considered, thus the spectrum of photospheric emission is taken to be a BB with single temperature. The diagnosis is easy to be performed for the emission with a hump-like spectrum. The hump-like spectrum could be well described with BB+NT model~\citep[e.g.][]{2013ApJ...770...32G},  where the NT component is denoted by a BAND or a exponential cut-off power law (CPL) function. However, there may be some problems in modeling with BB+CPL/BAND for QT-dominated spectra that are not hump-like, which are proved in the following analysis: 1) the obtained flux of BB component, $F_{\rm BB}$, is very small compared to the total flux of thermal component; 2) the low energy photon index $\alpha$ of CPL/BAND function may be still greater than the synchrotron death line~\citep{Preece_1998}, -2/3~\citep[e.g., as shown in Table 2 in][]{2022ApJ...932...25C}, which implies this so-called NT component may be also from photospheric emission; and 3) the obtained temperature of BB, $kT\sim1/3 E_{\rm p}$, and this means that there may be not two kinds of components (thermal and NT) in the emission.

In this paper, we perform the diagnosis for magnetization based on the modeling with mBB (or mBB+NT) model, to see if the conclusion is changed compared with that with BB+NT. Especially, a characteristic temperature with corresponding flux for the QT-dominated spectrum to replace the temperature and flux of a BB in BB+NT modeling in the diagnosis of magnetization. Two QT-dominated GRBs with known emission properties are taken as control samples in order to compare the two modelings. One control sample is GRB 210121A, of which the prompt emission is mainly from a typical pure hot fireball and the spectrum could be described with a mBB or NDP model~\citep{2021ApJ...922..237W,2022ApJ...931..112S}; the other one is GRB 210610B, which is also dominated by the photospheric emission and determined to be from a hybrid outflow based on fit results with the BB+NT model~\citep{2022ApJ...932...25C}. However, the spectrum of GRB 210610B is not evidently hump-like.

%However, quasi-thermal spectra are frequently observed. In some works~\cite[e.g.,][]{2022ApJ...932...25C}, the spectrum is described as a combined model of BB+BAND/CPL (CPL denotes exponential cut-off power law, BAND is the BAND function), and the obtained temperature ($kT$) and flux of the BB component are used to estimate $(1+\sigma_0)$ with `top-down' approach.%

GRB 221022B is recently detected by several missions, such as Fermi/GBM~\cite[GCN 32830,][]{2022GCN.32830....1P}, GRBAlpha~\cite[GCN 32844,][]{2021GCN.30697....1O} konus-wind~\cite[GCN 32864,][]{2022GCN.32109....1R} and \textit{Insight}-HXMT. The prompt emission has a long duration of about $50$ s. About one fourth GRBs have a beginning of photospheric emission (usually the first few pulses) and a general trend that pulses become softer over time with $\alpha$ becoming smaller~\citep[]{2021ApJS..254...35L}. The prompt emission of GRB 221022B behaves a typical evolution from thermal to non-thermal, which is a representative for some similar bursts. It is found that it may be from a hot fireball with an increasing Lorentz Factor at the beginning, which is well consistent with a typical expanding fireball scenario.

The paper is organized as follows: in Section~\ref{sec:TRana}, the methods for binning, background estimation, spectral fitting and model selection are introduced; in Section~\ref{sec:GAMP}, the characteristic temperature for modeling with mBB is introduced;
in Section~\ref{sec:topdownapproach}, the `top-down' approach and the hybrid model are introduced briefly; in Section~\ref{sec:test},
two control samples are used to make a comparison between the two modelings; in Section~\ref{sec:ana},  data analysis for GRB 221022B is performed; we discuss the emission mechanism of prompt emission, spectral and jet properties; the conclusion and summary are given in Section~\ref{sec:conclusion}. 

\section{methods for data analysis}\label{sec:TRana}
\subsection{Binning method of light curves for time-resolved spectra} \label{sec:Binning}
In this work, the Bayesian blocks (BBlocks) method introduced by \cite{2013ApJ...764..167S} and suggested by \cite{2014On}, is applied with a false alarm probability $p_{0}=0.01$ on light curves. In some cases, the blocks are coarse for fine time-resolved analysis. \cite{2014On} suggested that the constant cadence (CC) method is accurate when the cadence is not too coarse. Therefore, we take a combination of BBlocks and CC methods, fine binning of constant cadence are performed in each block, and only the bins with the signal-to-noise ratio (S$/$N)~$\geq$~40 at least in one detector should be utilized.
\subsection{Background estimation , spectral fitting method and model selection}

A polynomial is applied to fit all the energy channels and then interpolated into the signal interval to yield the background photon count estimate for GRB data. The Markov Chain Monte Carlo (MCMC) fitting is performed to find the parameters with the maximum Poisson likelihood. The best model is determined by the method of bayesian information criterion~\citep[BIC,][]{2016MNRAS.463.1144W},
where BIC is defined as
\begin{equation}\label{eq:BIC}
    \rm{BIC}=-2\rm{ln} \mathscr{L} +\rm{ln} (N)n,
\end{equation}
$\mathscr{L}$ denotes the likelihood
function for all these parameters based on a Bayesian prior, $\rm N$ is the number of data points, and $\rm n$ is the number of free parameters; $\Delta$BIC is the difference of BIC values of two models. A model that has a lower BIC value than the other is preferred. If the change of BIC between these two models, $\Delta$BIC is from 2 to 6, the preference for the model with the lower BIC is positive; if $\Delta$BIC is from 6 to 10, the preference
for that is strong; and if $\Delta$BIC is above 10, the preference is very strong.

%\section{ The multi-color blackbody modeling and the `top-down' approach for a hybrid jet model }
\section{The characteristic temperature and corresponding flux in modeling with mBB for quasi-thermal emission}\label{sec:GAMP}
We find many quasi-thermal emission could be described as a mBB function, in which the flux and the temperature of the individual Planck function is related by a power law with index $q$~\citep{2010ApJ...709L.172R},
\begin{equation}\label{eq:PL-mBB}
    F(T)=F_{\rm max} (T/T_{\rm max})^q,
\end{equation}
%\begin{equation}\label{eq:CPL-mBB}
%    F(T)=F_{\rm max} (T/T_{\rm c})^q %e^{-(T/T_{\rm c})^s},
%\end{equation} 
where $F_{\rm max}$ is the integrated flux of the Planck spectrum with a temperature of $T_{\rm max}$. For mBB, the spectrum consists of a superposition of Planck functions
in the temperature range from $T_{\rm min}$ to $T_{\rm max}$.

Here we assume that the multi-color superposition  is due to the angle dependence of the Doppler shift. Considering the connection between the observed temperature $T_{\rm obs}$ and $T^{\prime}_{\rm ph}$ in comoving frame, 
\begin{equation}
    T_{\rm obs}= D(\theta)T^{\prime}_{\rm ph}/(1+z),
\end{equation}
where $D(\theta)=(\Gamma(1-\beta \rm cos\theta))^{-1}$ is the Doppler factor, $\theta$ is the angle to the line of sight of the observer; $\Gamma$ is the Lorentz factor and $\beta=v/c$. For the emission from magnetized photosphere, $T^{\prime}_{\rm ph}$ is proportional to  $R_{\rm ph}(\theta)^{-1}$, $R_{\rm ph}(\theta)^{-(2+\delta)/3}$ or $R_{\rm ph}(\theta)^{-2/3}$ in different regimes ($\delta$ is the index of a power-law scaling for acceleration, $\Gamma\propto r^{\delta}$, $0<\delta\leq1/3$)~\cite[e.g.,][]{2015ApJ...801..103G}. With $R_{\rm ph}(\theta)\propto\frac{1}{\Gamma^2}+ \frac{\theta^2}{3}$~\citep{2008ApJ...682..463P}, and assuming that baryon loading is not very sensitive to $\theta$ within the opening angle, we easily get a direct conclusion that, the observed temperature reaches maximum in the direction of line of sight, which is the physical meaning of $T_{\rm max}$ in the observation with the simplest assumption of on-axis observation. Note that in fact density profiles of the outflow may be angle-dependent, and the emission probability is a function of $r$ with maximum at $r=R_{\rm ph}$, the case could be more complex. However, $T_{\rm max}$ could be taken as a probe for the outflow. We do not know the exact structure of the outflow or jet, thus, the following analysis is phenomenological and the spectrum is required to be well described with mBB function.

Considering the derivative of flux of mBB function has form of

\begin{equation}\label{eq:eq2}
%\begin{split}
\frac{dF(T)}{dT}=q\textbf{\\}\frac{F_{\rm max}}{T_{\rm max}^{q}}\textbf{\\} T^{q-1},
%F_{\rm mBB}(E, kT_{\rm min}, kT_{\rm max}, q, F_{\rm max})\\
%= \int^{kT_{\rm max}}_{\rm kT_{\rm min}} \frac{d}{dkT}( A(kT)\frac{E^3}{e^{E/kT}-1}) dkT,
%\end{split}
\end{equation}
integrating Equation~(\ref{eq:eq2}) to get a bolometric integrated spectral flux $F_{\rm mBB}$, we have
\begin{equation}
\label{eq:eqfmax}
F_{\rm max}= F_{\rm mBB}/(1-(\frac{kT_{\rm min}}{kT_{\rm max}})^{q}).
\end{equation}
$F_{\rm max}\gtrsim F_{\rm mBB} $ when $q>0$ and $kT_{\rm min}\ll kT_{\rm max}$. 
In some works~\citep[e.g.,][]{2018ApJ...866...13H}, other form of derivative of mBB function is used as 
\begin{equation}\label{eq:eq3}
\frac{dF(T)}{dT} =\frac{m+1}{(\frac{T_{\rm max}}{T_{\rm min}})^{m+1}-1}
\frac{F_{\rm mBB}}{T_{\rm min}}(\frac{T}{T_{\rm min}})^{m},  
\end{equation}
%where the normalization $A(kT)=\frac{15}{\pi^4}(kT)^{-4}F(kT, T_{\rm max}, q)$ and $F(kT, T_{\rm max}, q)$ has a form of Equation~(\ref{eq:PL-mBB}).
where $m$ is the index of the distribution of $dF/dT$, thus we have $q=m+1$. If we do not consider the intermediate photosphere~\citep[see][for similar details]{2022MNRAS.512.5693S} in the outflow, $kT_{\rm max}-F_{\rm max}$ could be taken as a good probe for the properties of the outflow.

\section{the `top-down' approach for a hybrid jet model}\label{sec:topdownapproach}
The work in~\cite{2015ApJ...801..103G} proposes an approach to \textbf{infer} the properties for the magnetized photosphere, such as for non-dissipative magnetized photosphere, we have
\begin{equation}\label{eq:sigma0p1}
    (1+\sigma_0)\propto(\frac{kT}{F_{\rm BB}})^{4/3}L_{\rm w}r_0^{2/3},
\end{equation}
where $L_{\rm w}$ is the entire luminosity of the outflow. Note that the estimation is based on the thermal component  which is described as a single blackbody function, without considering the structure of the outflow.

Here we give a simple derivative of the `top-down' approach~\citep{2015ApJ...801..103G} and an explanation of its physical origin. The acceleration of a GRB jet may be proceeded with two mechanisms: thermally driven or magnetically driven. %The former is relevant for a hot fireball, which proceeds very rapidly; whereas the latter is relevant for a Poynting flux dominated outflow, which proceeds relatively more slowly. For a more complicated hybrid jet system, we make the assumption that acceleration proceeds first thermally and then magnetically \citep[e.g.][]{1997ApJ...482L..29M,2003ApJ...596.1104V}.
Since thermal acceleration proceeds linearly, and the early magnetic acceleration below the magneto-sonic point also proceeds rapidly, we approximately assume that the ejecta first gets accelerated with $\Gamma \propto r$ until reaching a more generally defined \emph{rapid acceleration} radius $r_{\rm ra}$ defined by the larger one of the thermal coasting radius and the magneto-sonic point. Even though magnetic acceleration may deviate from the linear law below $r_{\rm ra}$, the mix with thermal acceleration would make the acceleration law in this phase very close to linear. Beyond $r_{\rm ra}$, the jet would undergo a relatively slow acceleration with $\Gamma \propto r^{\delta}$ until reaching a \emph{coasting radius} $r_{c}$.
if one ignores deceleration and energy loss,  the $\Gamma$ evolution for a hybrid system may be approximated as
\begin{eqnarray}
\label{eq:gammanodis}
\Gamma(r) = \left\{ \begin{array}{ll} \frac{r}{r_0}, & r_0<r<r_{\rm ra};\\
\Gamma_{\rm ra}\left(\frac{r}{r_{\rm ra}}\right)^{\delta}, &
 r_{\rm ra}<r<r_{\rm c}; \\
\Gamma_{\rm c}, &
r>r_{\rm c}, \\
\end{array} \right.
\end{eqnarray}

%we take the hybrid outflow without magnetic reconnection below the photosphere as a sample. %Note that we have $L_{\rm w}=L_{\rm m}+L_{\rm p}$ and $\eta(1+\sigma_0)=\frac{L_{\rm m}+L_{\rm p}}{\dot{M}c^2}$.
This scenario assumes that no magnetic field reconnection
occurs below the photosphere, so that no magnetic energy
is directly converted to particle energy and heat.
Without magnetic heating, the thermal energy undergoes adiabatic cooling, with $r^2e^{3/4}={\rm const}$ \citep[e.g.][]{1993MNRAS.263..861P}. Noticing $e \propto {T'}^4$ and the dynamical evolution Equation (\ref{eq:gammanodis}), one can derive the comoving temperature at the photosphere radius $r_{\rm ph}$ as
 \begin{eqnarray}
\label{eq:Tph1}
T'_{\rm ph} = \left\{ \begin{array}{ll} T_0\left(\frac{r_{\rm ph}}{r_0}\right)^{-1}, & r_0<r_{\rm ph}<r_{\rm ra};\\
T_0\left(\frac{r_{\rm ra}}{r_0}\right)^{-1}\left(\frac{r_{\rm ph}}{r_{\rm ra}}\right)^{-(2+\delta)/3}, &
r_{\rm ra}<r_{\rm ph}<r_{\rm c};  \\
T_0\left(\frac{r_{\rm ra}}{r_0}\right)^{-1}\left(\frac{r_{\rm c}}{r_{\rm ra}}\right)^{-(2+\delta)/3}\left(\frac{r_{\rm ph}}{r_{\rm c}}\right)^{-2/3}, &
r_{\rm ph}>r_{\rm c}. \\
\end{array} \right.
\end{eqnarray}
Here
\begin{equation}
T_0 \simeq \left(\frac{L_{\rm w}}{4\pi r_0^2 a c (1+\sigma_0)}\right)^{1/4},
\end{equation}
is the temperature at $r_0$, $a=7.56 \times 10^{-15} {\rm erg~cm^{-3}~K^{-4}}$ is radiation density constant. Given the central engine parameters $L_{\rm w}$, $r_0$, $\eta$ and $\sigma_0$, we can derive all the relevant photosphere properties with equations from (\ref{eq:L_sigma0}), (\ref{eq:gammanodis}) and (\ref{eq:Tph1}), as long as the slow magnetic acceleration index $\delta$ is determined.  The largest $\delta$ is 1/3, which is achievable for an impulsive, non-dissipative magnetic shell \citep{2011MNRAS.411.1323G}. For different central engine parameters, $\Gamma_{\rm ra}$ can have two possible values: $\eta$ or $[\eta (1+\sigma_0)]^{1/3}$. For each case, the photosphere radius $r_{\rm ph}$ can be in three different regimes separated by $r_{\rm ra}$ and $r_c$, thus there are six saturated regimes of rapid acceleration, including Regime I: $r_{\rm ph}<r_{\rm ra}$ with $\eta>(1+\sigma_0)^{1/2}$; 
    Regime II: $r_{\rm ra}<r_{\rm ph}<r_{\rm c}$ with $\eta>(1+\sigma_0)^{1/2}$; Regime III: $r_{\rm ph}>r_{\rm c}$ with $\eta>(1+\sigma_0)^{1/2}$; Regime V: $r_{\rm ra}<r_{\rm ph}<r_{\rm c}$ with $\eta<(1+\sigma_0)^{1/2}$; Regime VI: $r_{\rm ph}>r_{\rm c}$ with $\eta<(1+\sigma_0)^{1/2}$; and Regime IV: ($r_{\rm ph}<r_{\rm ra}$ with $\eta<(1+\sigma_0)^{1/2}$). Relevant parameters are derived in each regime, including $r_{\rm ra}$ and $r_{\rm c}$, along with the photosphere properties, i.e. $r_{\rm ph}$, $(1+\sigma_0)$, $kT_{\rm ob}$, and $F_{\rm BB}$ accordingly, and  more details of the formulae could be found in ~\cite{2015ApJ...801..103G}.

%For the unsaturated regimes, such as Regime I ($r_{\rm ph}<r_{\rm ra}$ with $\eta>(1+\sigma_0)^{1/2}$), IV ($r_{\rm ph}<r_{\rm ra}$ with $\eta<(1+\sigma_0)^{1/2}$), $\eta$ can not be determined from the observation, which are not considered.

%Some QT spectra might be broader than a BB, and it is more reasonable that their spectra are described with a mBB function rather than a BB function with a single temperature. Thus, we should use the parameters obtained from modeling with mBB to estimate the properties of photospheric emission.
In the modeling with BB+NT, $kT$ and $F_{\rm BB}$ is used in the `top-down' approach to diagnose the jet properties,  while for the modeling with mBB, we use $kT_{\rm max}$ and $F_{\rm max}$ as the characteristic temperature and the corresponding flux for the QT emission, to replace $kT$ and $F_{\rm BB}$ . If $(1+\sigma_0)>1$, QT emissions are from magnetized photosphere. $\sigma_{15}$ determined at radius $r=10^{15}$ cm is used to speculate the origin of NT emissions if it exists in the emission. Note that we do not know the initial radius $r_0$ for the outflow, thus we first assume a series of $r_0=10^7 - 10^9$ cm in the calculation, however, it is found that only a narrow range of $r_0$ could give reasonable results with constraints from $\eta$, $(1+\sigma_0)$,  $r_{\rm ra}$, $r_{\rm ph}$, and $r_{\rm c}$ for each regime. We consider these saturated regimes of rapid acceleration, including Regime II, III, V and VI. For the unsaturated regimes, such as Regimes I and IV, $\eta$ can not be determined from the observation, which are not considered. Besides, the typical value, $Y=2$, is considered in the procedure, where $Y$ is the ratio between the total fireball energy and the energy emitted, and $Y\gtrsim1$. Constant or decreasing $r_0$ values with time are both accepted, if
one considers the depletion of the envelope.

If $(1+\sigma_0)\simeq1$,  it is similar to the case of the pure hot fireball. In this case, $r_0$ could be directly determined by the method in \cite{2007ApJ...664L...1P}. The generalized method with $kT_{\rm max}\mbox{--}F_{\rm max}$ can be applied in this case as well. 

\section{Test with control samples: GRB 210121A and GRB 210610B}\label{sec:test}
\subsection{A control sample of hybrid outflow: GRB 210610B}\label{sec:210610B} 
 As analyzed in \citep{2022ApJ...932...25C}, GRB 210610B has low-energy indices  that are ranging from -0.2 to -0.5 and all greater than the
synchrotron cutoff, which implies that most of the emission are photospheric; the top-down approach is performed based on the modeling with a BB+NT to diagnose the magnetization, and it suggests that GRB 210610B is originated from a hybrid jet. Note that there exist observations of the afterglow~\citep{2021GCN.30275....1K}, which implies that it is not from a pure hot fireball.
In this section, GRB 210610B is used to test this method based on the modeling with mBB+NT, and $(1+\sigma_0)>1$ is expected at least in one time bin in time-resolved analysis. 

\begin{deluxetable*}{lccccccccccc}
\tabletypesize{\tiny}
%\tabletypesize{\small}
\tablewidth{0pt}
\tablecaption{The time-resolved results of GRB 210610B with mBB (+NT). The preferred model is labeled by a underline. The ranges of  estimated $r_0$ of mBB modeling are in \textit{italics}. \label{tab:fitresTR_210610B}}
\tablehead{ 
\colhead{Time bins}
&\colhead{model}
&\colhead{$m$}
&\colhead{$kT_{\rm min}$}
&\colhead{$kT_{\rm max}$} 
&\colhead{$F_T$} 
&\colhead{$\alpha$} 
&\colhead{$F_{NT}$}
&\colhead{BIC} &\colhead{$\frac{\chi^2}{ndof}$}
&\colhead{$r_0$ ($\times10^{8}$ cm)}
&\colhead{$r_0$ ($\times10^{8}$ cm)}
\\
\colhead{(s)} 
&\colhead{} 
&\colhead{} 
&\colhead{(keV)}
&\colhead{(keV)}
&\colhead{(10$^{-6}$ erg cm$^{-2}$ s$^{-1}$)}
&\colhead{}
&\colhead{(10$^{-6}$ erg cm$^{-2}$ s$^{-1}$)} &\colhead{}
&\colhead{}%\\\hline
&\colhead{Regime II}
&\colhead{Regime III}
}
\startdata
[-10, 25] & mBB &0.62$^{+0.14}_{-0.18}$ &4.2$^{+2.5}_{-2.9}$ &118.2$^{+18.8}_{-12.6}$ &0.51$^{+0.04}_{-0.06}$ &  &  &185.7 &$\frac{176.7}{168}$ & \textit{[1.1, 2.6]} & \textit{[0.3, 0.5]} \\
&\underline{mBB+PL} &0.80$^{+0.10}_{-0.20}$ &1.7$^{+3.8}_{-1.0}$ &133.0$^{+31.3}_{-36.7}$ &0.30$^{+0.07}_{-0.10}$ &-1.72$^{+0.32}_{-0.31}$  &0.16$^{+0.03}_{-0.06}$ &179.6 &$\frac{166.2}{166}$ &[0.5, 1.6] &[0.1, 0.3]\\\hline
[25, 35] &mBB
&0.34$^{+0.04}_{-0.02}$ &5.1$^{+0.3}_{-2.4}$ &167.2$^{+8.1}_{-7.6}$ &5.27$^{+0.06}_{-0.13}$ &  &  &178.1 &$\frac{169.1}{168}$& \textit{[1.8, 4.3]}  & \textit{[0.4, 0.7]} \\
&\underline{mBB+PL}&0.26$^{+0.04}_{-0.08}$ &9.1$^{+1.8}_{-1.8}$ &176.0$^{+8.7}_{-7.8}$ &5.37$^{+0.13}_{-0.13}$ &-2.66$^{+0.39}_{-0.22}$  &0.23$^{+0.07}_{-0.06}$ &171.5 &$\frac{158.1}{166}$ &[1.7, 3.9]&[0.3, 0.6]\\\hline
[35, 45] &\underline{mBB} &0.29$^{+0.03}_{-0.05}$ &3.1$^{+1.2}_{-1.1}$ &125.6$^{+8.0}_{-5.5}$ &2.81$^{+0.09}_{-0.07}$ &  &  &159.9 &$\frac{150.9}{168}$& \textit{[2.4, 5.5]}& \textit{[0.4, 0.8]}\\
&mBB+PL &-0.98$^{+0.15}_{-1.85}$ &25.7$^{+4.0}_{-18.2}$ &453.6$^{+246.6}_{-327.8}$ &2.58$^{+0.29}_{-0.25}$ &-1.63$^{+0.06}_{-0.22}$ &1.75$^{+0.61}_{-0.30}$ &300.2 &$\frac{286.8}{166}$ &  & \\\hline
[45, 55] &mBB
&-0.90$^{+0.08}_{-0.04}$ &9.5$^{+0.8}_{-4.5}$ &415.4$^{+320.0}_{-127.2}$ &1.65$^{+0.04}_{-0.04}$ &  &  &189.7 &$\frac{180.8}{168}$&  &\\
&\underline{mBB+PL} &-0.46$^{+0.32}_{-0.27}$ &9.6$^{+1.9}_{-1.7}$ &102.9$^{+22.3}_{-33.5}$ &1.13$^{+0.07}_{-0.09}$ &-2.56$^{+0.14}_{-0.56}$ &0.25$^{+0.05}_{-0.09}$ &155.6 &$\frac{142.2}{166}$&[3.0, 6.1]&[0.5, 0.9]\\\hline
[55, 100] &mBB &-1.21$^{+0.91}_{-0.15}$ &6.0$^{+0.7}_{-3.0}$ &875.0$^{+230.8}_{-182.2}$ &0.26$^{+0.05}_{-0.08}$ &   &  &287.4 &$\frac{278.5}{168}$&  &\\
&\underline{mBB+PL} &-0.23$^{+0.25}_{-0.27}$ &3.7$^{+1.6}_{-1.3}$ &45.7$^{+5.8}_{-6.5}$ &0.19$^{+0.02}_{-0.01}$ &-2.53$^{+0.05}_{-0.32}$ &0.10$^{+0.01}_{-0.03}$ &238.1 &$\frac{224.7}{166}$ &[4.0, 11.6]&[0.5, 1.1]\\\hline
\enddata
\end{deluxetable*}

\begin{figure*}
\begin{center}
\includegraphics[width=0.3\textwidth]{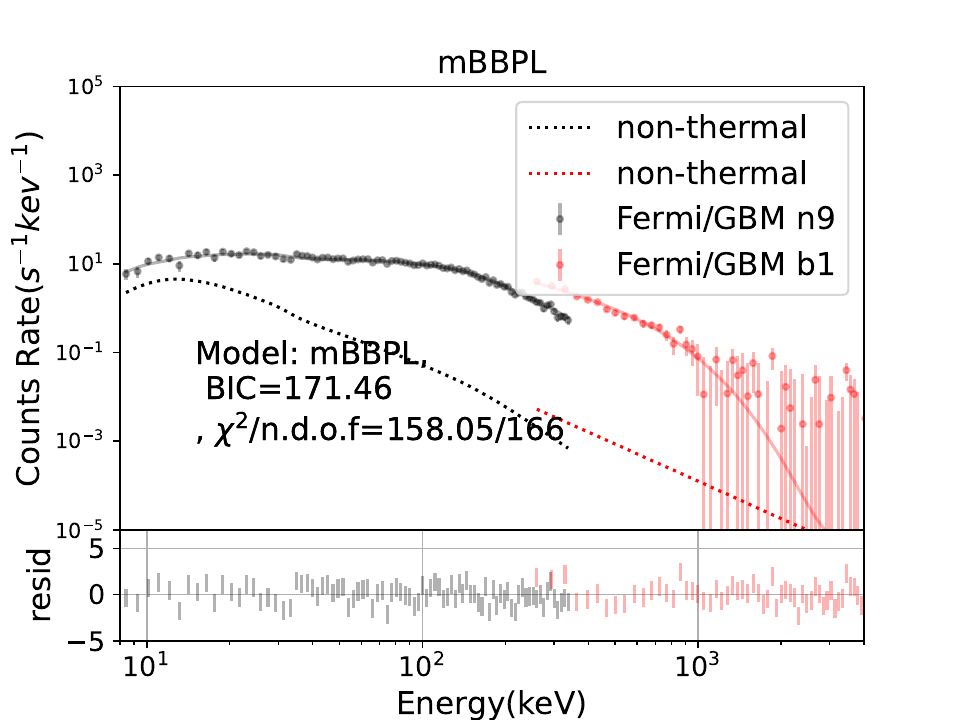}\put(-130,90){(a){$T_0+[25, 35]$ s } }
\includegraphics[width=0.3\textwidth]{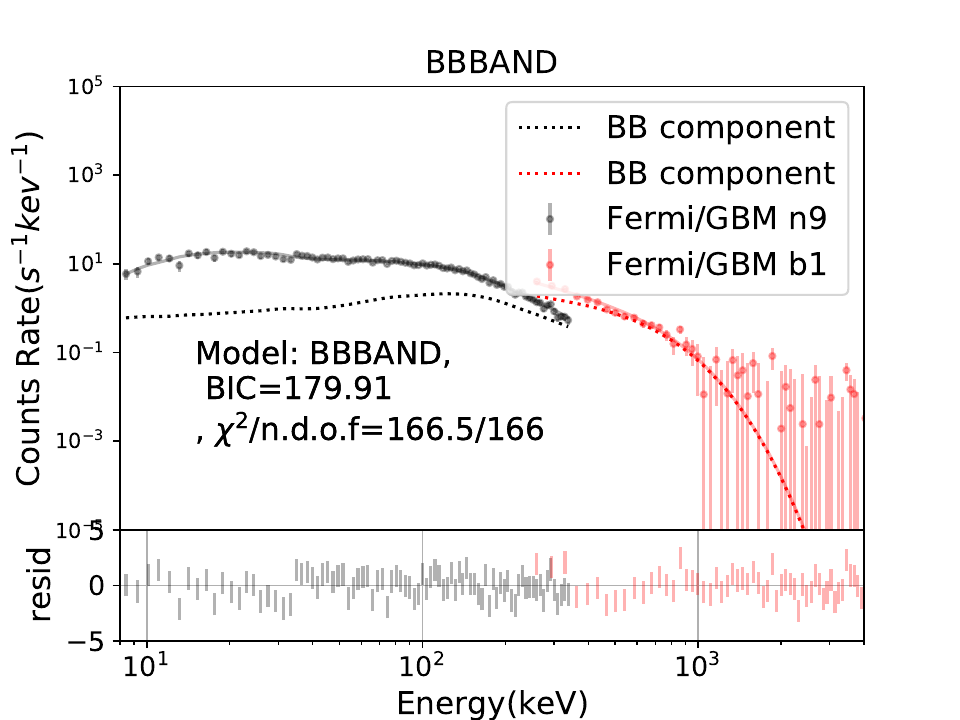}\put(-110,90){(b) }
\includegraphics[width=0.3\textwidth]{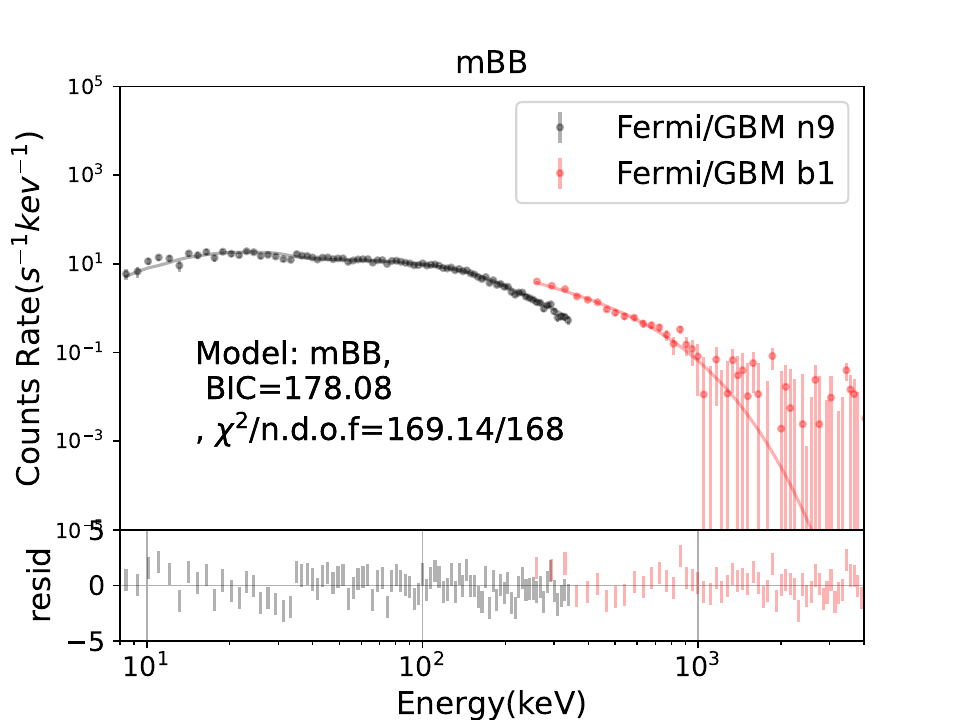}\put(-110,90){(c) }
\\
\includegraphics[width=0.3\textwidth]{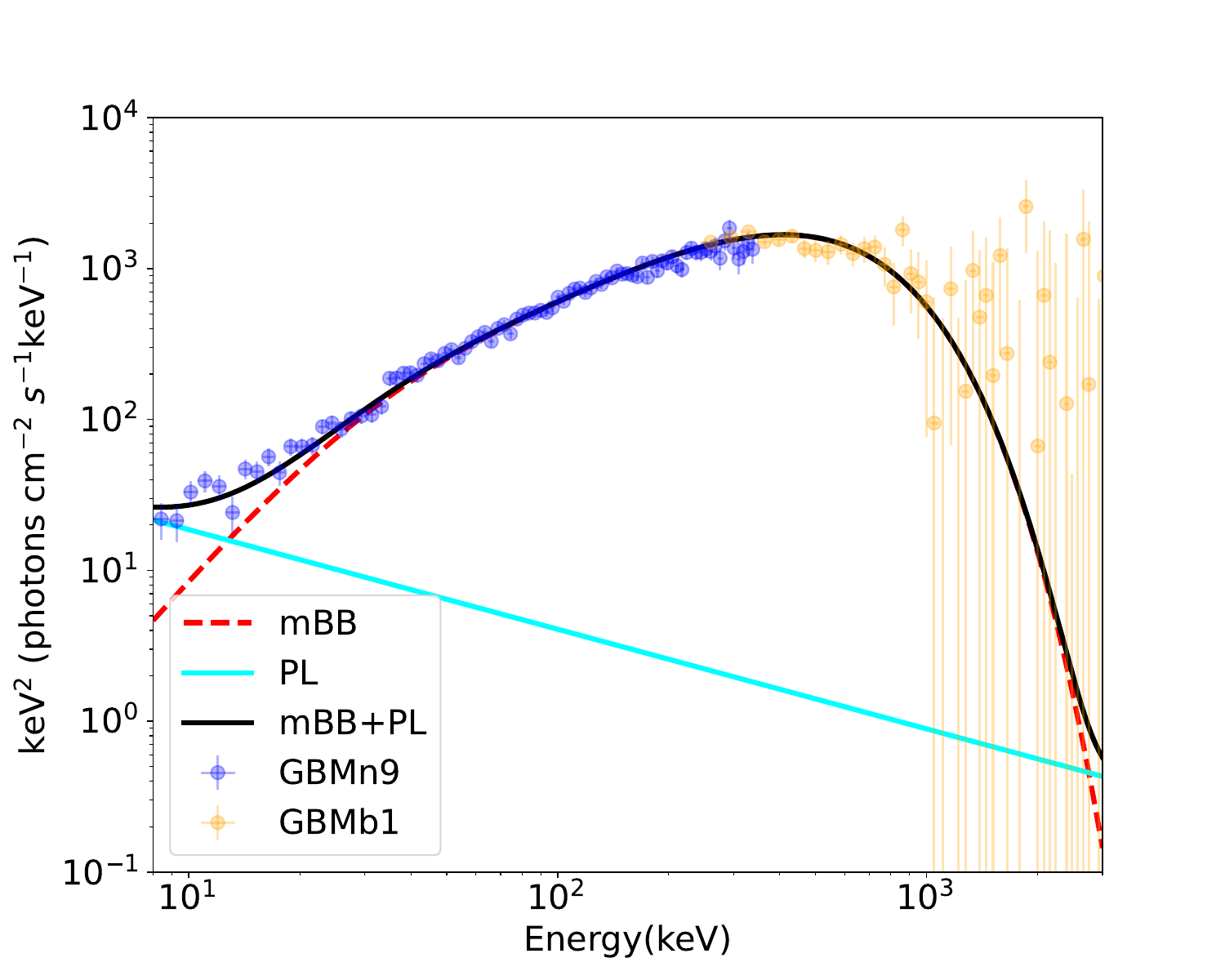}\put(-130,90){(d)} 
\includegraphics[width=0.3\textwidth]{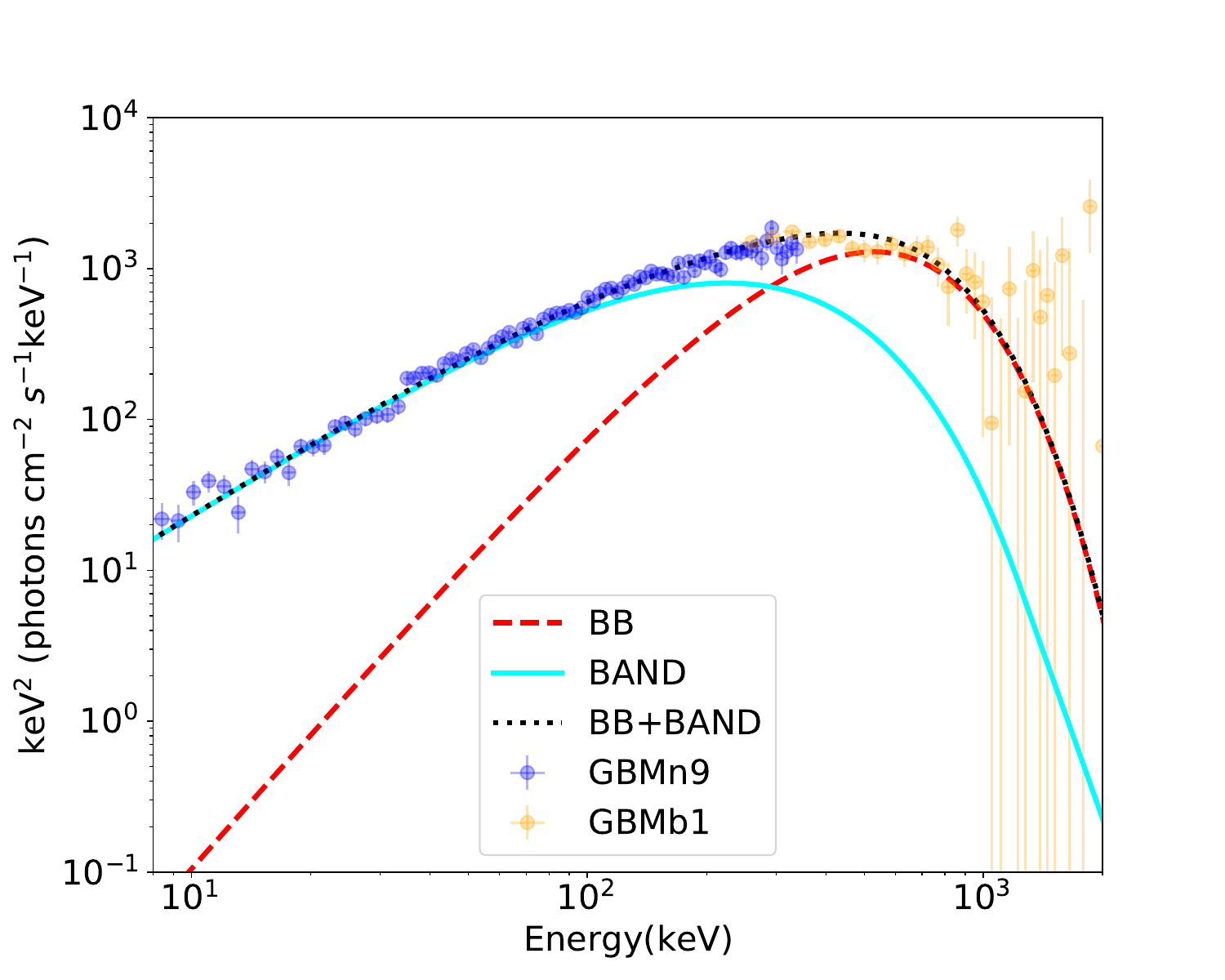}\put(-110,90){(e) }
\includegraphics[width=0.3\textwidth]{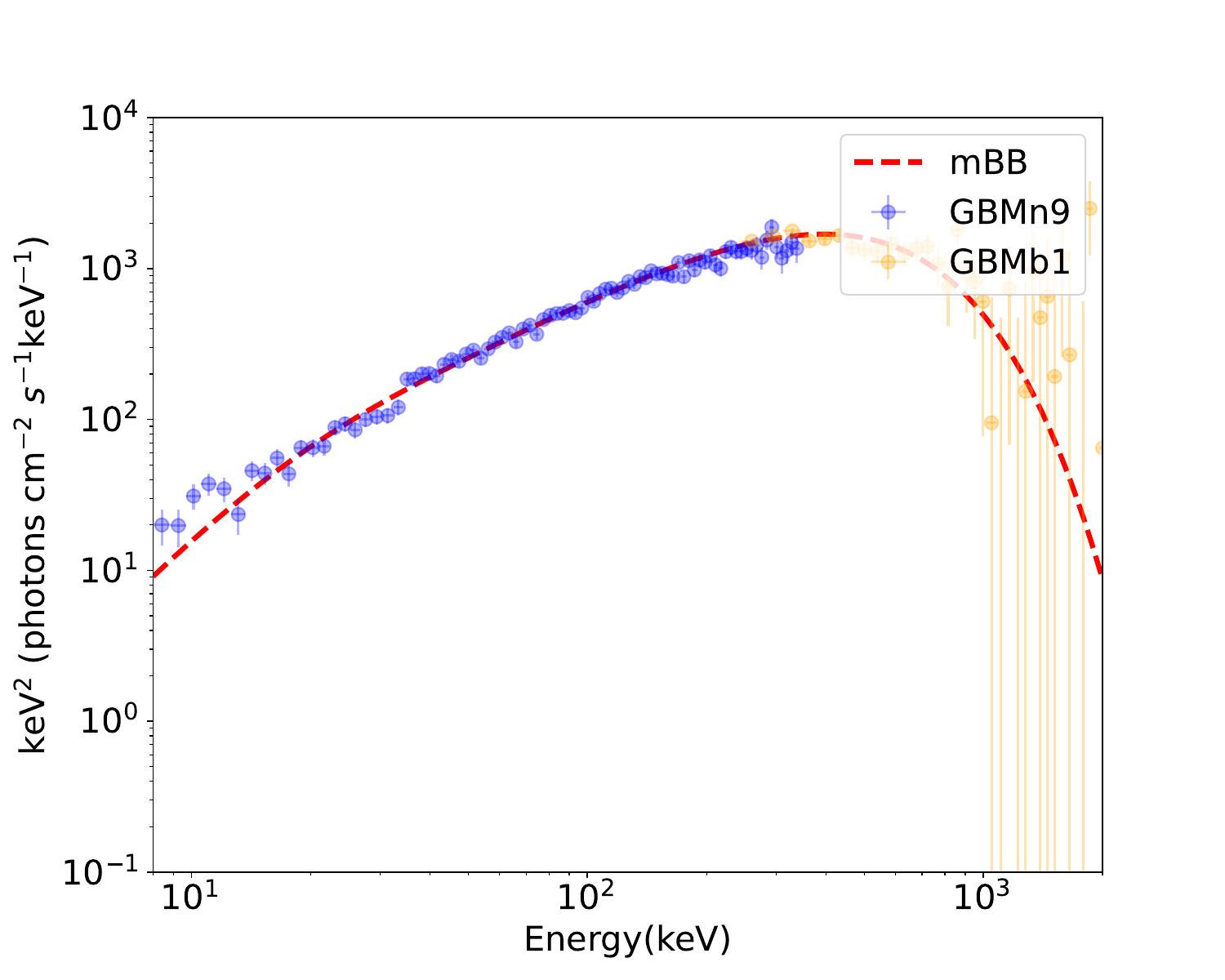}\put(-110,90){(f)  }\\

 \includegraphics[width=0.32\textwidth]{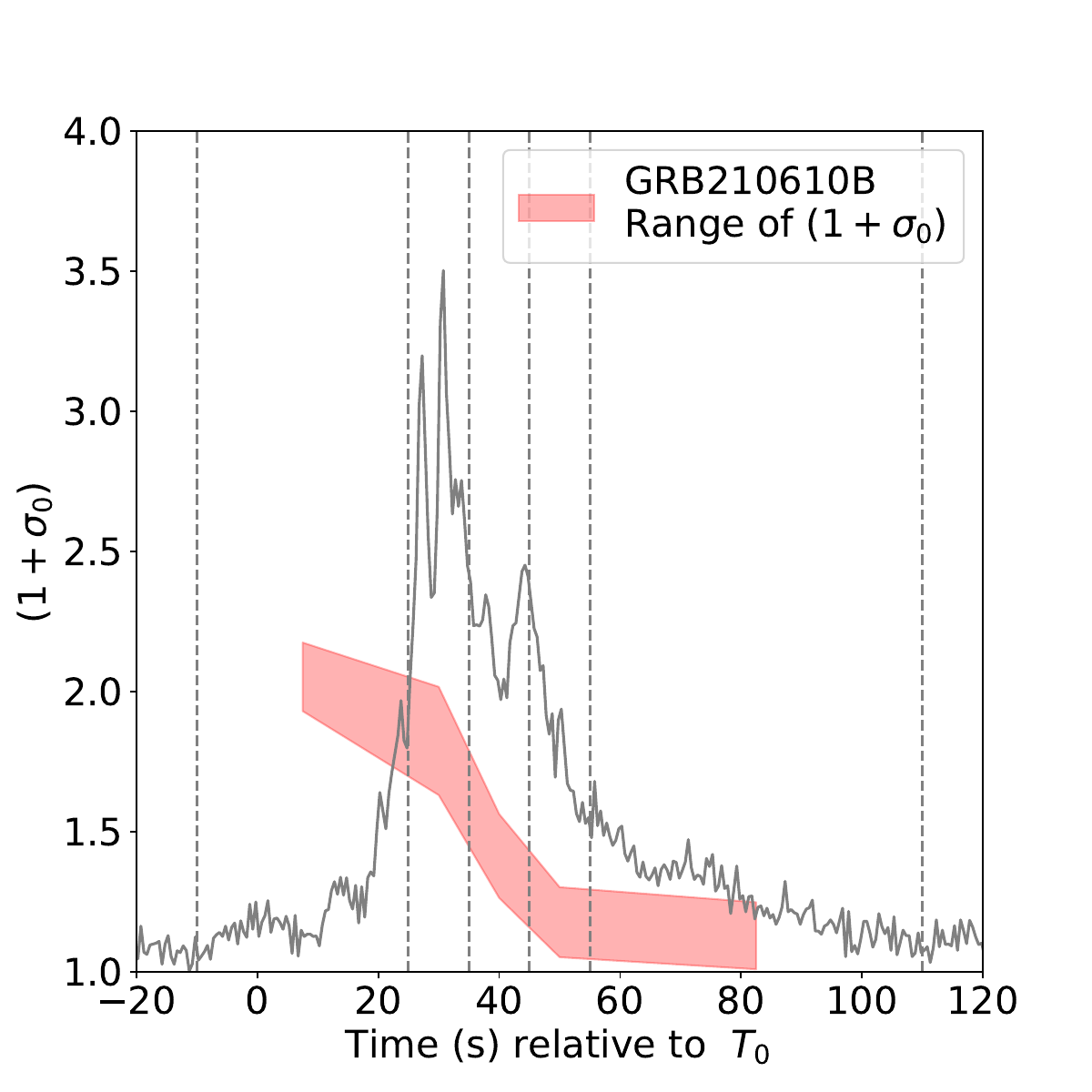}\put(-140,110){(g) with best modeling}
  \includegraphics[width=0.32\textwidth]{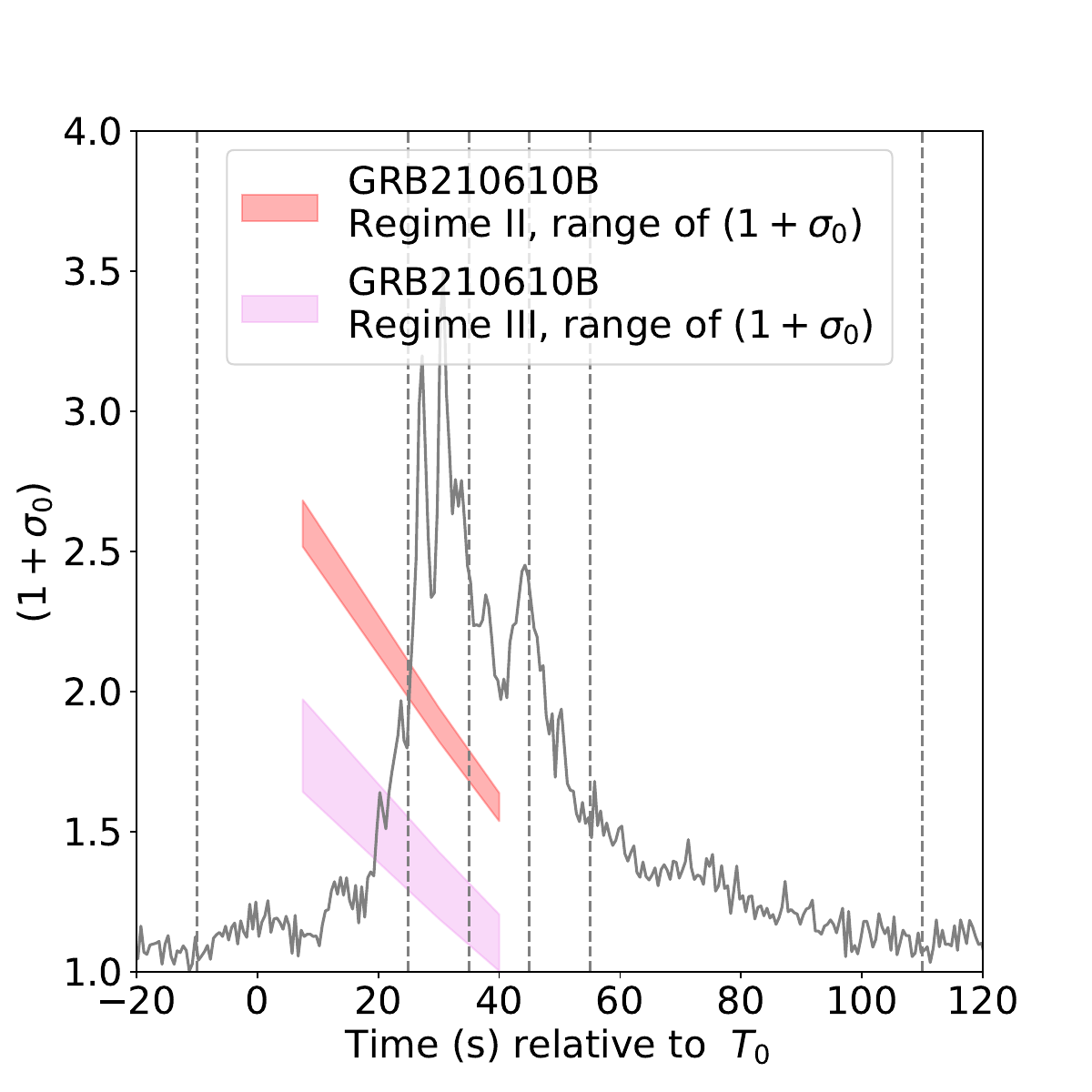}\put(-140,100){(h) with a single mBB modeling}
 \includegraphics[width=0.36\textwidth]{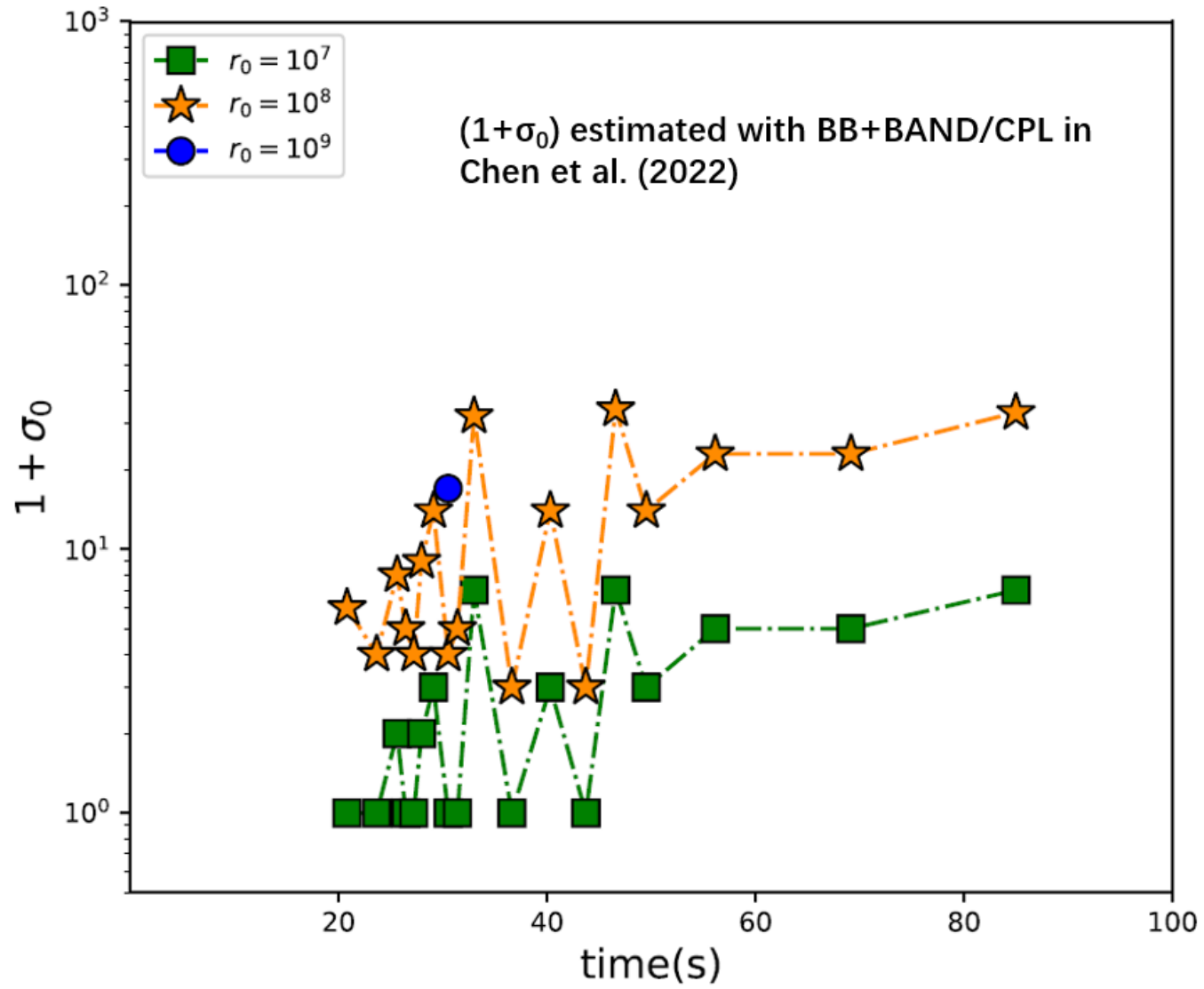}\put(-140,100){(i)}
  \\
 \caption{\label{fig:magnetization_controlsamples2} (a) , (b) and (c) are spectra in $[25, 35]$ s and fit results with mBB+PL, BB+BAND and mBB models, while the corresponding panels below, (d), (e) and (f) are model shapes in the $E^2N(E)$ form. (g) and (h) $1+\sigma_0$ estimated from the best modeling and a single mBB modeling. The gray line denotes shape of the normalized light curve, the same below. The vertical dashed lines denote the time internals. (i) $1+\sigma_0$ estimated from modeling with BB+NT, adapted from Figure 3 (b) in \cite{2022ApJ...932...25C}.}
 \end{center}
\end{figure*}

 We divide the prompt phase into five time intervals: $[-10, 25]$ s, $[25, 35]$ s, $[35,45]$ s, $[45,55]$ s and $[55, 110]$ s relative to $T_0$ ($T_0$ is the trigger time of the corresponding GRB, the same below). The preferred model for the spectrum in $[35,45]$ s is mBB while the other time-resolved spectra are best described by a combined model of mBB+PL. The fit results with the best models are shown in Table~\ref{tab:fitresTR_210610B}. We take the spectrum in $T_0+[25, 35]$ s which has the highest flux for example, to show the modeling selection between BB+BAND and mBB+PL. As shown in Figure~\ref{fig:magnetization_controlsamples2} (a)-(d), the spectrum is not hump-like. If modeling with a CPL function, one has $\alpha=-0.37\pm0.02$ and $E_{\rm p}=379.5\pm3.5$ keV. The extracted parameters and BIC values for these two combined models are listed as below:
 \begin{itemize}
     \item BB+BAND: for BB, $kT=135.3\pm14.1$ keV and flux $F_{\rm BB}=(2.81\pm0.04)\times10^{-6}$ erg cm$^{-2}$ s$^{-1}$; for BAND: $\alpha=-0.35\pm0.02$, $\beta=-9.37\pm0.72$, $E_{\rm p}=224.5\pm10.6$ keV and $F_{\rm BAND}=2.63\pm0.27$ erg cm$^{-2}$ s$^{-1}$. BIC=179.9 with a freedom degree of 166;~\footnote{In fact, there may exist some uncertainties in modeling with BB+NT in these cases of NOT hump-like spectra, the extracted BB component may have a $E_{\rm p}$ consistent with the whole spectrum and a comparable flux with the total flux, or have a much lower $E_{\rm p}$ and flux as that in Figure~\ref{fig:GRB221022B_m2_20} (d).}
     \item mBB+PL: for mBB, $m=0.26\pm0.06$, $kT_{\rm min}=9.1\pm1.1$ keV, $kT_{\rm max}=176.0^{+8.7}_{-7.8}$ keV, $F_{\rm mBB}=(5.37\pm0.13)\times10^{-6}$ erg cm$^{-2}$,  s$^{-1}$; for PL: $\alpha=-2.66^{+0.39}_{-0.22}$, $F_{\rm PL}=0.23\pm0.06$ erg cm$^{-2}$ s$^{-1}$. BIC=171.5 with a freedom degree of 166.
 \end{itemize}

  These two models both describe the spectrum well, however, from BIC values, mBB model is preferred with $\Delta$BIC= 8.4. Besides, there are some features of extracted parameters of BB+BAND which could not be ignored: 1) $\alpha\sim-0.35$ is well above -2/3 for the BAND component in BB+BAND, which may be difficult to be interpreted as a NT emission; %for comparison, $\alpha \sim-1$ in the BAND component in the modeling with BB+BAND in GRB 110721A~\citep{2013MNRAS.433.2739I};
  2) $kT=135.3\pm14.1$ keV for BB components corresponds to a peak energy of $3kT\sim400$ keV, which is consistent with $E_{\rm p}$ of the whole spectrum. Considering that the component described with BAND is also QT, we inferred that the QT component is dominant. %For comparison, the peak ($\sim100$ keV) of the BB component is far from $E_{\rm p}$ ($\sim$MeV) of the BAND component in BB+BAND modeling in GRB 110721A. 
  Therefore, mBB+PL is better from  both of the physical meaning and model selection.
 
 The ranges of $r_0$ are estimated and listed in the last two columns in Table~\ref{tab:fitresTR_210610B}. $r_0$ ranges around $10^{8}$ cm in the Regimes II and III, while there exists no reasonable $r_0$ values from $10^{6}-10^{10}$ cm to satisfy Regimes V and VI. Note that $r_0$ values in each regime have a roughly increasing trend with time. With assuming that $r_0$ is constant (or has a decreasing trend)~\footnote{If
one considers the depletion of the envelope, $r_0$ should decrease
with time.} during the prompt phase, 
 the possible regime for $T_0+[-10, 25]$ s is II, while Regime III is reasonable for the other part of prompt phase.  A narrow range ($0.5\times10^{8}- 0.6\times10^{8}$ cm) is obtained.

 The estimated $(1+\sigma_0)$ is greater than 1 for most bins as shown in Figure~\ref{fig:magnetization_controlsamples2} (e). The maximum $1+\sigma_0$ is about 2, which means a moderate magnetization in the outflow. Thus the outflow for GRB 210610B is diagnosed to be hybrid, which is consistent with the conclusion based on the modeling with BB+NT in ~\cite{2022ApJ...932...25C}. The estimated $\sigma_{15}$ is less than 1, which corresponds to a coasting regime at these radii, and the NT emission is mainly from IS mechanism. This is also consistent with the conclusion in \cite{2022ApJ...932...25C}. Note that the contribution from BAND component covers nearly a half in BB+BAND (see the fit results in [35, 45] s listed above)\footnote{ For other time bins, the contribution from BAND component covers even much more, as shown in Table 2 in \cite{2022ApJ...932...25C}.}; however, IS mechanism has a small radiative efficiency~\citep[only a few percent,][]{2011ApJ...726...90Z}, thus it seems not very reasonable that this `NT' component denoted by a BAND function is the NT emission produced from the IS mechanism. For mBB+PL model, the NT emission has a much smaller flux, which provides more reasonable explanation.

 Figure~\ref{fig:magnetization_controlsamples2} (f) shows $(1+\sigma_0)$ estimated with BB+BAND modeling and different $r_0$ values of 10$^{7}$ cm, $10^{8}$ cm and $10^{9}$ cm, in which $(1+\sigma_0)$ ranges from 1 to several tens~\citep{2022ApJ...932...25C}. We note that the maximum $(1+\sigma_0)>10$ of $r_0=10^8$ cm, which is larger than that from mBB model. In fact, if $\sigma_0=\frac{L_{\rm p}}{L_{\rm m}}>10$, the outflow is dominated by the Poynting flux, a NT emission could be expected due to the magnetic dissipation, such as internal-collision-induced magnetic reconnection and turbulence ~\citep[ICMART,][]{2011ApJ...726...90Z}. Some other Poynting-flux-dominated GRBs, e.g. GRB 211211A~\citep{2023ApJ...943..146C} and GRB 221009A~\citep{2023ApJ...947L..11Y}, have  prompt phases dominated by NT emissions, which are greatly different from the observation of GRB 210610B. Thus we infer that a large $\sigma_0$ may not be reasonable for GRB 210610B. Note that in Equation~(\ref{eq:sigma0p1}), the denominator, $F_{\rm BB}$ from the modeling with BB+NT is smaller than the truth, which may account for the large $\sigma_0$ obtained with BB+BAND model.
 
 In our analysis, PL denotes the possible NT emission caused by IS mechanism. However, the uncertainty of the modeling and the impact on the diagnosis should be considered. Thus whether impacts from the lack of a NT component affects the conclusion is tested with a single mBB modeling. The modeling with a single mBB is shown in Table~\ref{tab:fitresTR_210610B} and Figure~\ref{fig:magnetization_controlsamples2} (c) and (f).  A similar procedure is performed, and the magnetization is estimated, as shown in Figure~\ref{fig:magnetization_controlsamples2} (h)\footnote{We note that after $T_0$+45 s, $m$ of mBB values is nearly -1, or even less that -1, which means that the spectral shape is far from that dominated by thermal component. Thus we do not use the parameters of a single mBB after $T_0$+45 s}. It is found that there exists at least one time bin with $(1+\sigma_0)>1$ estimated with a single mBB, namely, the conclusion is similar with that of mBB+NT qualitatively. This could be understood that the PL contribution is small even if an additional PL is needed in the modeling, and the parameters of mBB in two modeling are similar.  Thus, we infer that the lack of NT component does not affect the effectiveness of this method much.
 
 The possibility of sub-photosphere magnetic dissipation should be discussed. In this scenario, the magnetic energy is dissipated below the photosphere via e.g. magnetic reconnection and could be converted to the thermal energy at least partially. Thus an effective temperature can be derived, which would be the temperature if the emission is fully thermalized~\citep[e.g.,][]{2000ApJ...529..146E,2007ApJ...666.1012T}. Practically, it would serve as an estimate of the lower limit of $E_{\rm p}$ of a magnetically dissipative photosphere emission. The photosphere emission properties could be obtained under the assumption of significant magnetic dissipation. However, because GRB 210610B has a lower $E_{\rm p}$ below 1 MeV, the estimated magnetization will not be large even if we perform this estimation, which is similar to that under assumption of magnetized photosphere without sub-photosphere magnetization. Moreover, since the proportion of the magnetic energy converting to heat during the sub-photosphere dissipation is unknown to us, the estimated ranges for parameters ($\eta$, $r_0$, $\sigma_0$) could be even larger with considering the possible regimes, which seem not meaningful estimations for outflow properties. In fact, in some similar papers in which the hybrid model and `top-down' approach are used~\citep[e.g.,][]{2020ApJ...894..100L,2022ApJ...932...25C}, the case of sub-photosphere magnetic dissipation is also not considered for similar reasons. 
  
 \subsection{A control sample of a pure hot fireball: GRB 210121A }
 As discussed in \cite{2021ApJ...922..237W} and \cite{2022ApJ...931..112S}, the spectrum of prompt emission of GRB 210121A is well consistent with that of NDP model, which implies that it is mainly from a pure hot fireball. There is no report about the existence of its afterglow. The spectra in both epochs ([$T_0-0.01$, $T_0+2.19$] s and [$T_0+2.8$, $T_0+14.8$] s, $T_0$ is the trigger time of the corresponding GRB, the same below) are well described with a NDP model of a pure hot fireball. $(1+\sigma_0)\simeq1$ is expected for both epochs. 
 
\begin{deluxetable*}{lcccccccc}
%\tabletypesize{\tiny}
\tabletypesize{\small}
\tablewidth{0pt}
\tablecaption{The time-resolved results of GRB 210121A with the mBB.\label{tab:fitresTR_210121A}}
\tablehead{ 
\colhead{Time bins}
&\colhead{$m$}
&\colhead{$kT_{\rm min}$}
&\colhead{$kT_{\rm max}$} 
&\colhead{$F_T$} 
&\colhead{BIC} &\colhead{$\frac{\chi^2}{ndof}$}
&\colhead{$r_0$ ($\times10^{8}$ cm)%\tablenotemark{a}
}
&\colhead{$r_0$ ($\times10^{7}$ cm)}\\
\colhead{(s)} 
&\colhead{} 
&\colhead{(keV)}
&\colhead{(keV)}
&\colhead{(10$^{-6}$ erg cm$^{-2}$ s$^{-1}$)}
&\colhead{}
&\colhead{}
&\colhead{Regime II}
&\colhead{Regime III}
%\\\hline
}
\startdata
 [-0.01, 2.19]
 &0.54$^{+0.05}_{-0.05}$ &3.0$^{+2.5}_{-1.5}$ &397.0$^{+28.7}_{-23.5}$ &16.46$^{+0.91}_{-0.71}$
 &206.4 
 &$\frac{197.3}{192}$ &[0.4, 0.8] &[0.3, 2.3]\\\hline
 [2.80, 14.90] &0.33$^{+0.05}_{-0.04}$ &1.8$^{+0.9}_{-0.6}$ &356.5$^{+39.8}_{-28.1}$ &4.75$^{+0.31}_{-0.28}$   &202.8 &$\frac{193.6}{192}$&[0.2, 0.5]&[0.1, 1.1]\\\hline
\enddata
%\tablenotetext{a}{Here the ranges of $r_0$ are given with $Y=2$. %Considering that GRB 210121A is from a pure hot fireball in fact and $Y=1$, the possible ranges of $r_0$ are [0.6, 0.8]$\times10^8$ cm , [0.4, 0.5]$\times10^8$ cm for two time intervals in Regimes II, and [0.4, 0.7]$\times10^8$ cm, [0.1, 0.3]$\times10^8$ cm in Regime III. 
%}
\end{deluxetable*}
 
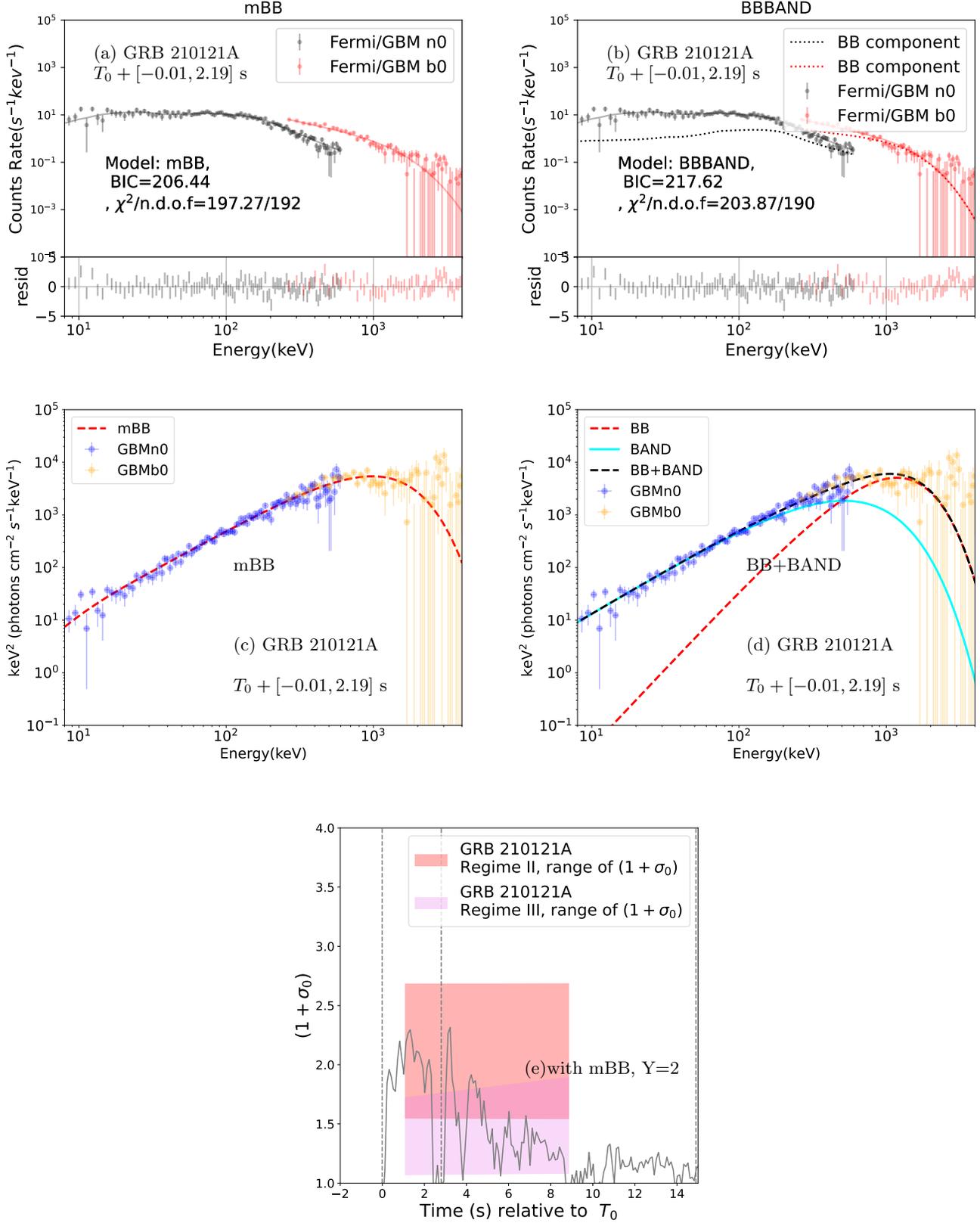
\begin{figure*}
\begin{center}
 \centering
 \includegraphics[width=0.5\textwidth]{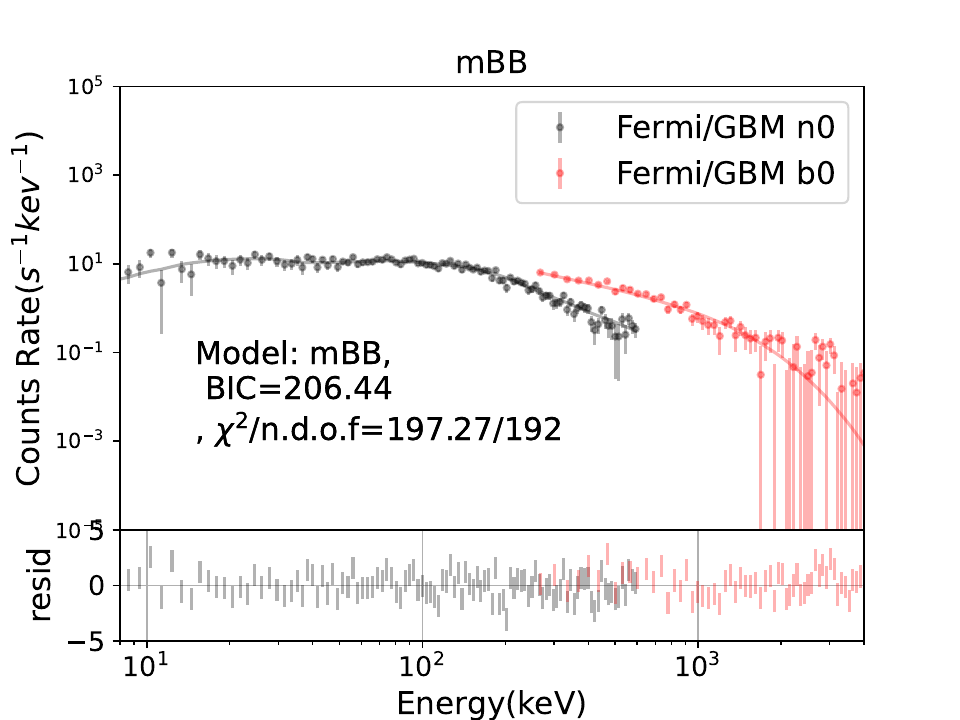}\put(-210,150){(a) GRB 210121A }\put(-210,140){$T_0+[-0.01, 2.19]$ s }
   \includegraphics[width=0.5\textwidth]{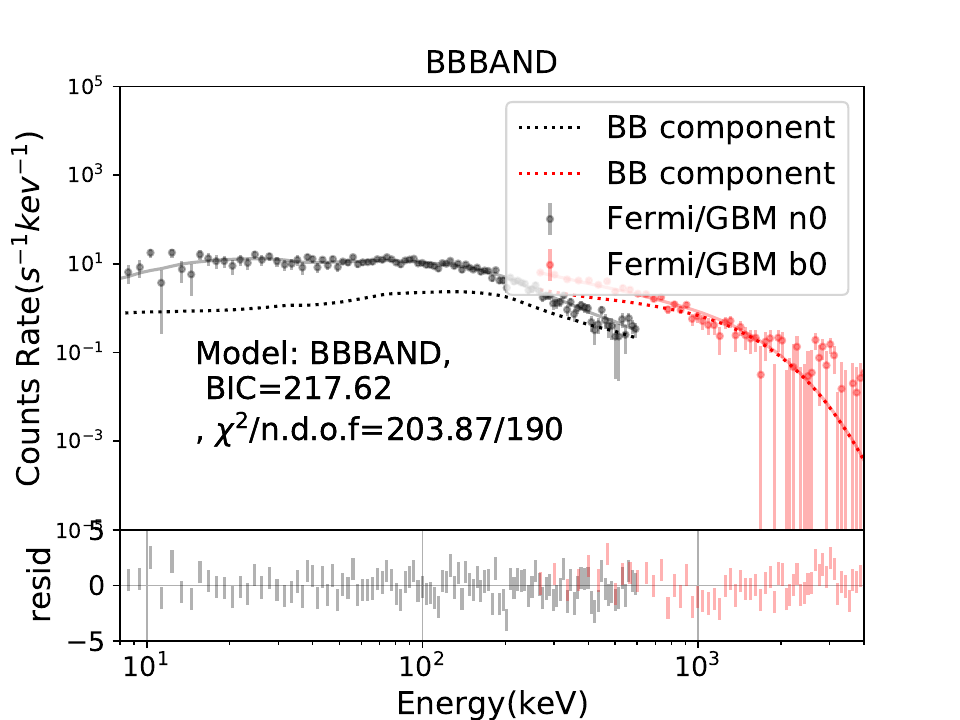}\put(-210,150){(b) GRB 210121A }\put(-210,140){$T_0+[-0.01, 2.19]$ s }\\
   \includegraphics[width=0.5\textwidth]{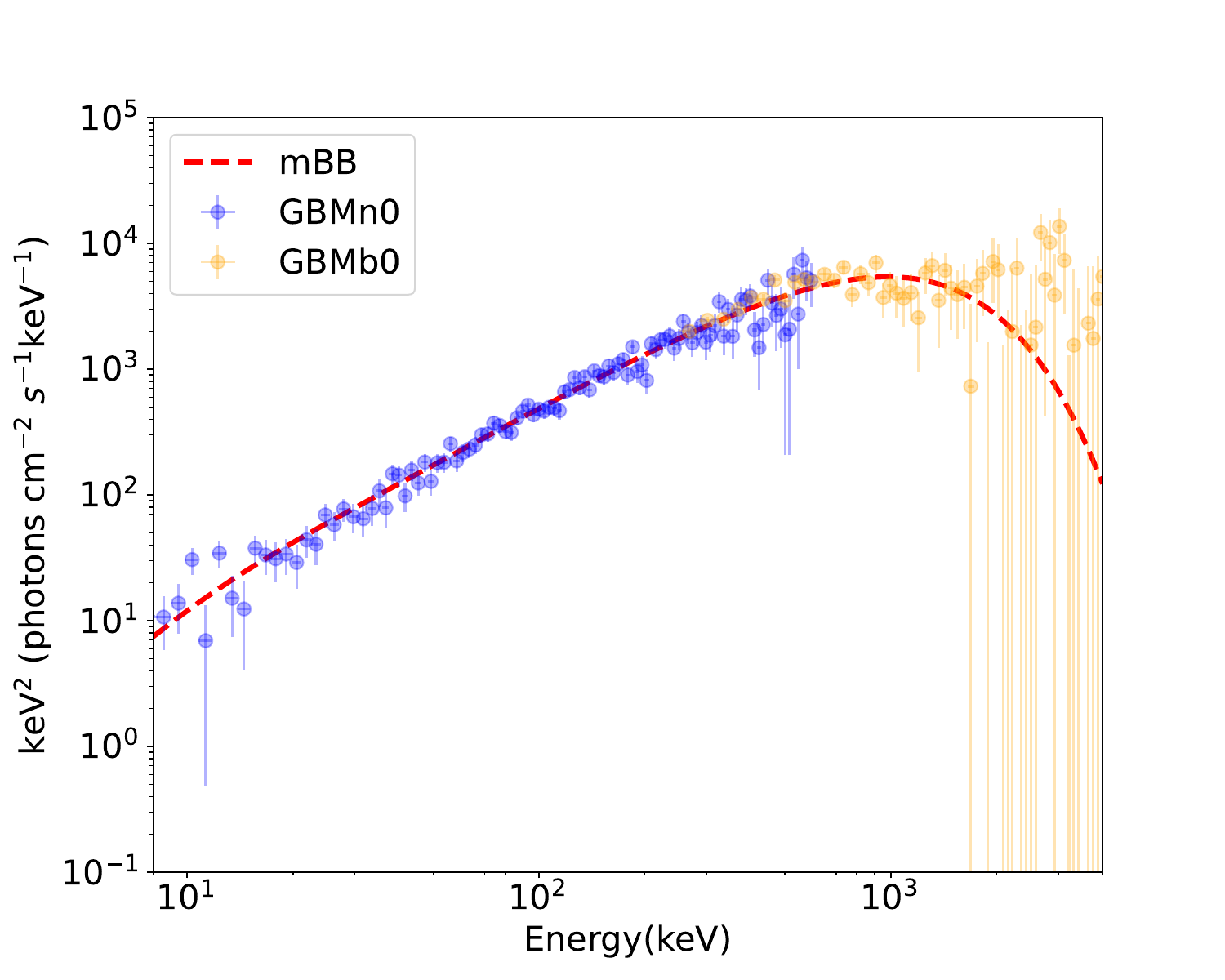}\put(-140,60){(c) GRB 210121A }\put(-140,40){$T_0+[-0.01, 2.19]$ s} \put(-140,100){mBB}
   \includegraphics[width=0.5\textwidth]{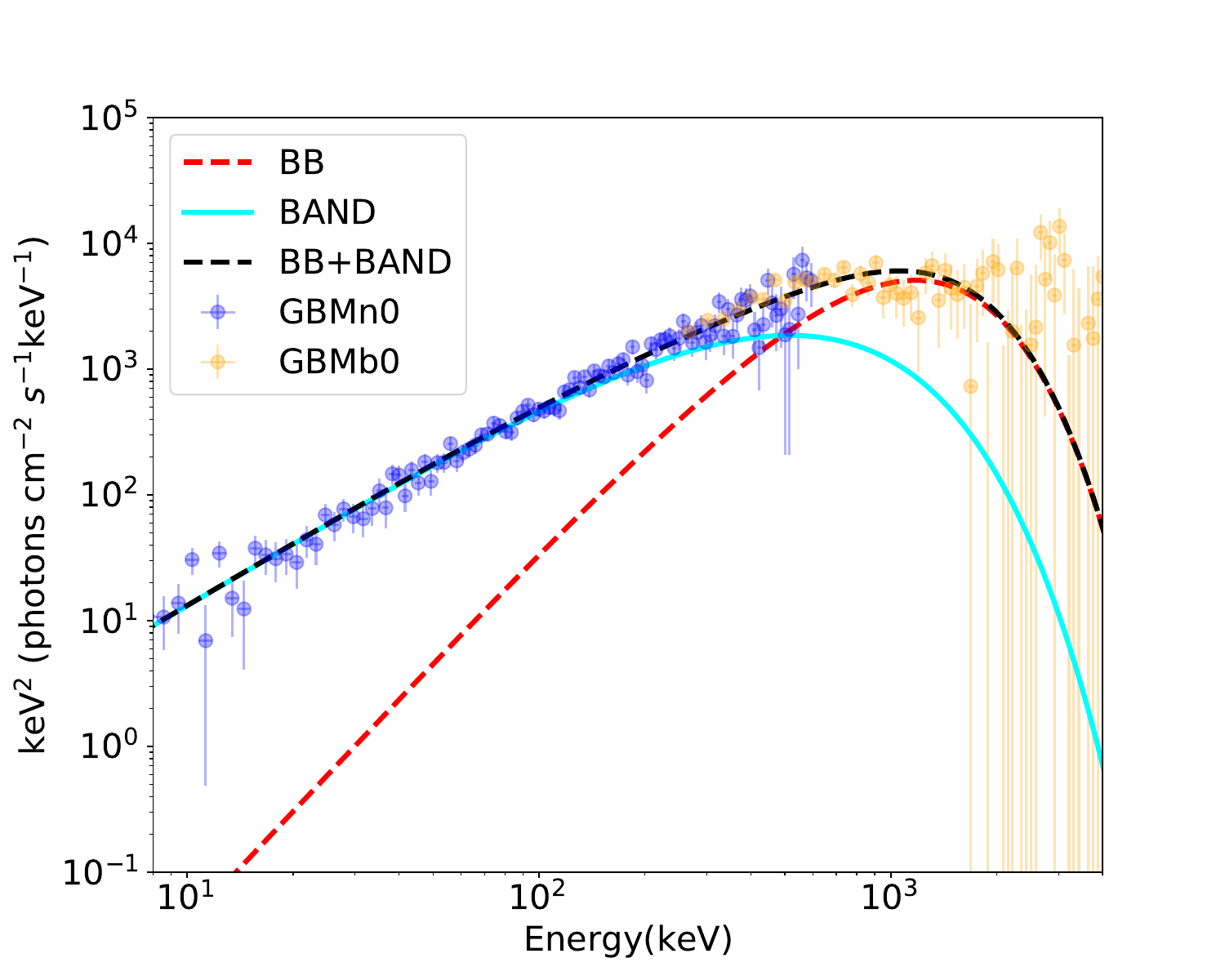}\put(-140,60){(d) GRB 210121A }\put(-140,40){$T_0+[-0.01, 2.19]$ s} \put(-140,100){BB+BAND}\\
     \includegraphics[width=0.45\textwidth]{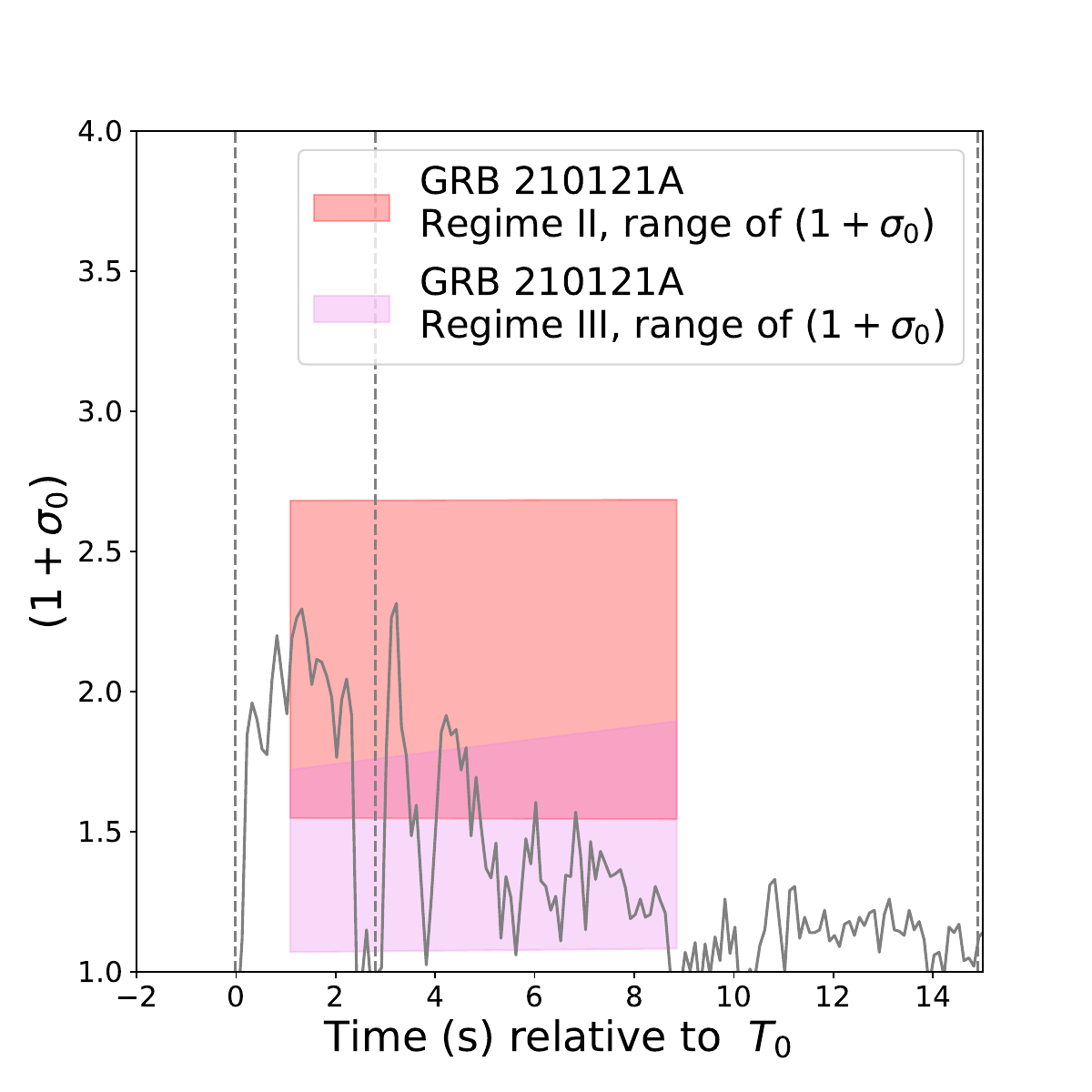}\put(-110,80){(e)with mBB, Y=2}
   \caption{\label{fig:magnetization_controlsamples} (a)-(b) are the spectra and fit results with mBB and BB+BAND models and BIC values while the corresponding panels below, (c) and (d) are model shapes in the $E^2N(E)$ form. (e) The ranges of $(1+\sigma_0)$ for GRB 210121A with $Y=2$. %Note that they are both estimated with the fit results from mBB modeling.
   }
\end{center}
\end{figure*}

Here we take [$T_0-0.01$, $T_0+2.19$] s with the highest flux for example to compare the difference between modeling with mBB and BB+BAND models. The fit results and spectra are shown in Figure~\ref{fig:magnetization_controlsamples} (a)-(d), and parameters are listed as below,
\begin{itemize}
     \item BB+BAND: for BB, $kT=300.3\pm25.1$ keV and $F_{\rm BB}=(11.07\pm0.14)\times10^{-6}$ erg cm$^{-2}$ s$^{-1}$; for BAND: $\alpha=-0.32\pm0.04$, $\beta=-9.92\pm0.92$, $E_{\rm p}=224.5\pm10.6$ keV and $F_{\rm BAND}=6.06\pm0.19$ erg cm$^{-2}$ s$^{-1}$. BIC=217.2 with a freedom degree of 190;
     \item mBB: for mBB, $m=0.54\pm0.05$, $kT_{\rm min}=3.04^{+2.5}_{-1.5}$ keV, $kT_{\rm max}=397.0^{+28.7}_{-23.5}$ keV, $F_{\rm mBB}=(16.46^{+0.91}_{-0.71})\times10^{-6}$ erg cm$^{-2}$,  s$^{-1}$; BIC=197.3 with a freedom degree of 192.
 \end{itemize}

It seems that the case is similar to that in GRB 210610A, mBB is still preferred in GRB 210121A. For GRB 210121A, spectra are well described with a mBB model for both epochs, and the fit results are listed in Table~\ref{tab:fitresTR_210121A}. %More details about the modeling parameters could be found in the published works~\cite[e.g.][]{2021ApJ...922..237W,2022arXiv220409430S}, thus we do not show them again in this analysis. 

We find both of two regimes (II and III) are possible. The possible $r_0$ ranges are listed in Table~\ref{tab:fitresTR_210121A}. In the same regime, the estimated $r_0$ values could share ranges and satisfy the assumption of a constant or decreasing trend in $r_0$ with time, which are different from those in GRB 210610B shown in Table~\ref{tab:fitresTR_210610B}. As shown in Figure~\ref{fig:magnetization_controlsamples} (e), though $1+\sigma_0$ could be greater than 1 in Regime II, $\sigma_0\lesssim0.1$ and $1+\sigma_0\simeq1$ is NOT ruled out for both epochs, because Regime III could work as well. For comparison, in GRB 210610B, $1+\sigma_0\simeq1$ is ruled out at least in one time bin. Therefore, diagnosis based on mBB modeling supports that GRB 210121A is from a pure hot fireball, or has a very low magnetization in the outflow.  

There are not former works about GRB 210121A based on modeling with BB+NT. It could be inferred that $\sigma_0$ might be overestimated due to the underestimated $F_{\rm BB}$, as discussed in the case of GRB 210610B. However, we note that even if based on mBB modeling, the estimation for $(1+\sigma_0)$ with `top-down' approach covers a wide range (from 1.1 to 2.5) as shown in Figure~\ref{fig:magnetization_controlsamples} (e), which implies that it may be not sensitive enough to distinguish the pure hot fireball from a hybrid outflow with a mild magnetization.

 \subsection{Comparison between two modelings and implications}

 From analyses of the control samples, a summary is obtained from the comparison between mBB model (mBB or mBB+NT: $kT_{\rm max}-F_{\rm max}$) and BB+NT model (BB+NT: $kT_{\rm BB}-F_{\rm BB}$):
 \begin{itemize}
     \item{Modeling:} 
     
     mBB model could be preferred from both of the physical meaning and model selection, if the QT spectrum is not hump-like;
     \item{The estimation for magnetization:}
     
     For GRB 210610B, the estimated $1+\sigma_0$ with mBB model seems more reasonable compared with that with BB+NT model, though the qualitative conclusions of the jet properties from two models for the diagnosis are similar. 
     \item{Can it distinguish pure hot fireball from hybrid outflow? }
     
     For GRB 210121A,
     $1+\sigma_0\simeq1$ is not ruled out, while for GRB 210610B, $1+\sigma_0\simeq1$ is ruled out at least in one time bin. It seems that whether $1+\sigma_0\simeq1$ is ruled out could serve as a possible criterion to distinguish the pure hot fireball from the hybrid jet. It is inferred that if modeling with BB+NT, $1+\sigma_0$ might be overestimated, which causes an erroneous judgement. However,  considering the wide range of estimated $1+\sigma_0$, the uncertainties in the data analysis, and the differences between the models and real data, we think it may be difficult to distinguish between these two cases only from this criterion, especially in the case of a mild magnetization.
 \end{itemize}

\section{spectral and jet properties of GRB 221022B}\label{sec:ana}
\subsection{spectral properties of GRB 221022B }
\begin{figure*}
\begin{center}
 \centering
  \includegraphics[width=0.5\textwidth]{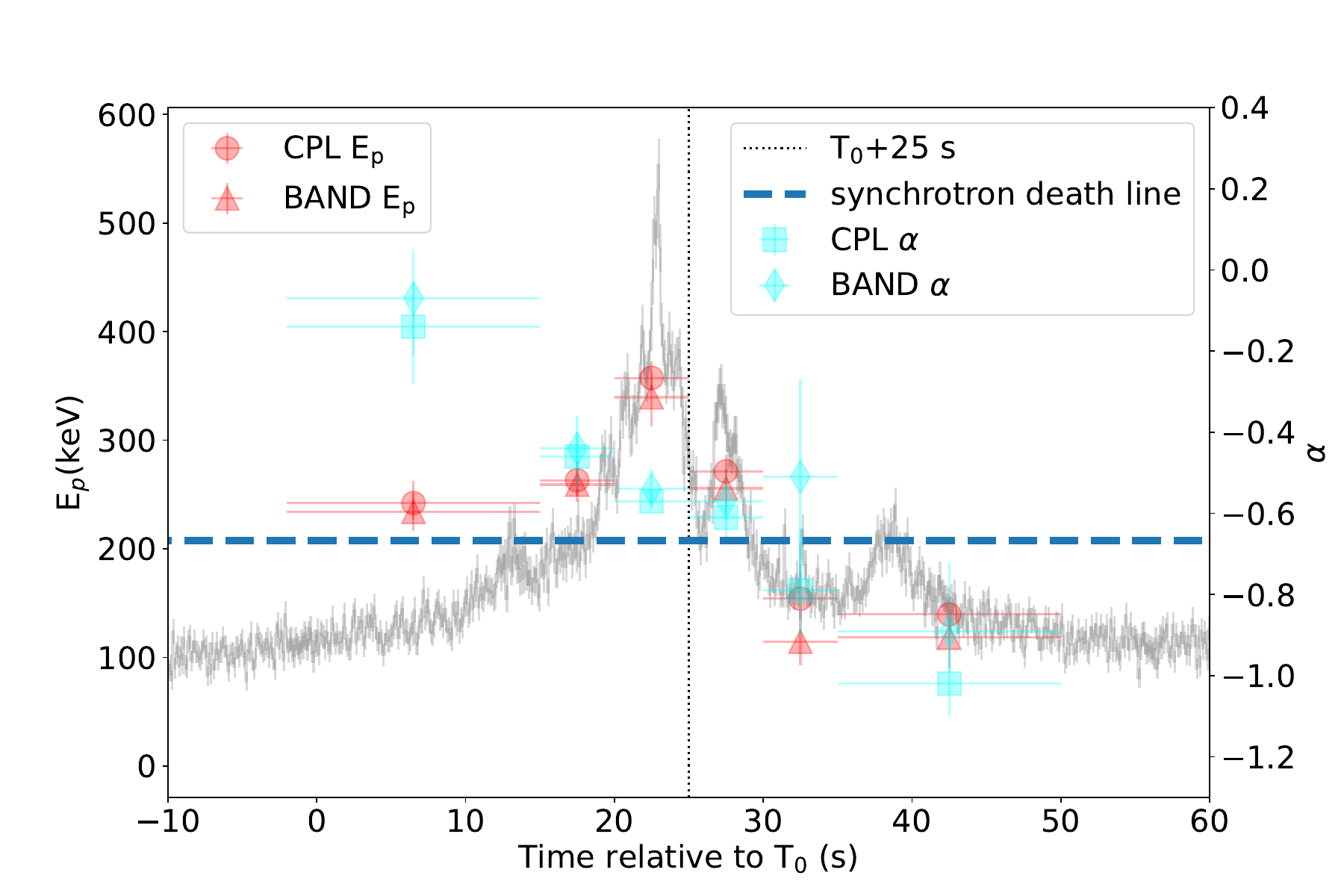}\put(-110,100){(a) GRB 221022B}
  \includegraphics[width=0.49\textwidth]{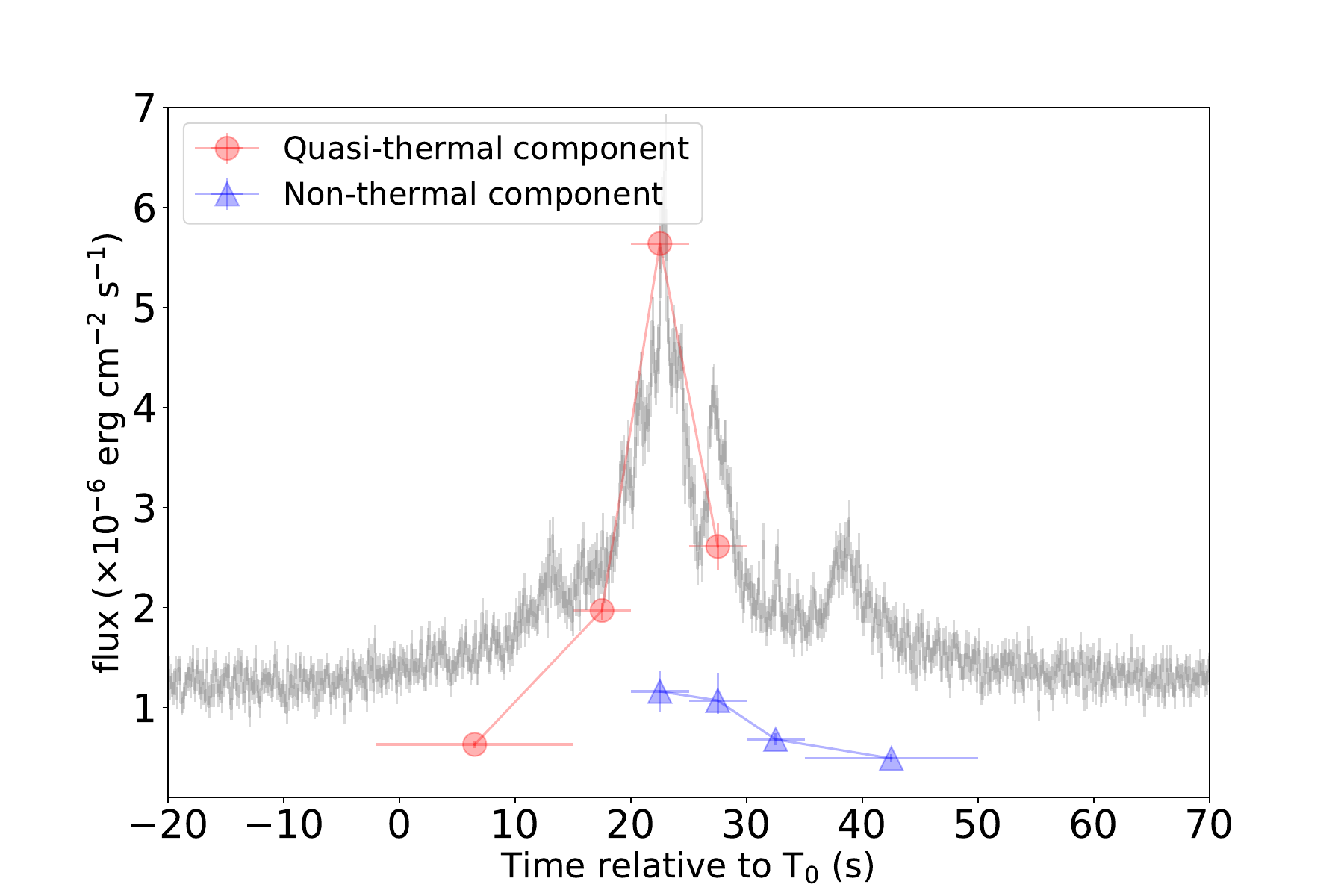}\put(-110,100){(b) GRB 221022B}
\caption{\label{fig:LC_GRB221022B}(a) The evolution of $\alpha$ and $E_{\rm p}$ of GRB 221022B. (b) The flux of QT and NT components of GRB 221022B. }
\end{center}
\end{figure*}

GRB 221022B is a long burst with duration of $\sim 50$ s. The light curves from Fermi-GBM and HXMT are shown in Figure.~\ref{fig:LC_GRB221022B}. The data are from one brightest NaI detector (NaI 4) and one brightest BGO detector (BGO 0) of Fermi/GBM. We check the data from $[T_0-300, T_0+300]$ s in different energy bands and find that there exists no peaking or bump structure in the background. Fittings with BAND, CPL and mBB are performed on the time-integrated spectrum from $T_0-2$ s to $T_0+50$ s which contains $95\%$ photons. 
$E_{\rm p}$ and bolometric flux in [$T_0-2$, $T_0$ + 50] s are determined to be 274.4$^{+15.0}_{-10.7}$ keV and 1.55$^{+0.03}_{-0.05}\times10^{-6}$ erg cm$^{-2}$ s$^{-1}$ with BAND function. Amati-relation and observed long GRBs with known redshift~\cite[e.g.,][]{2018NatCo...9..447Z} give a measurement of $E_{\rm p}$-$E_{\rm iso}$ correlation as log$E_{\mathrm{p}}(1+z)=(2.22\pm0.03)+(0.47\pm0.03)$log$(E_{\mathrm{iso}, \gamma}/10^{52})$. The clustering of long bursts gives a broad range of $z=0.61^{+1.49}_{-0.23}$ for this burst, and in the following analysis we take $z=0.61$. 

%Constant cadence~\citep[CC,][]{2014On} method and Bayesian blocks~\citep[BBlocks,][]{2013ApJ...764..167S} method with a false alarm probability $p_0$= 0.01 are used for binning in time-resolved analysis. We also require the signal-to-noise ratio (S$/$N)~$\geq$40 at least in one detector, so we combine some adjacent bins in the beginning and end of the burst. The time bins are [-2, 15] s, [15, 20] s, [20, 25] s, [25, 30] s, [30, 35] s and [35, 50] s relative to $T_0$.

 The fit results with BAND, CPL, mBB and mBB+PL functions are listed in Table.~\ref{tab:fitresTR}. Spectra of all the time bins could be described well with empirical models (Band and CPL) as well as mBB models. The evolution of $\alpha$ and $E_{\rm p}$ values are shown in Figure.~\ref{fig:LC_GRB221022B} (a). $\alpha$ decreases with time generally, while $E_{\rm p}$ tracks the flux roughly. For the first four bins from $T_0-2$ s to 30 s it is well above the death line of synchrotron, while after $T_0+30$ s, it gradually drops to -1, a typical value for GRBs.  
 
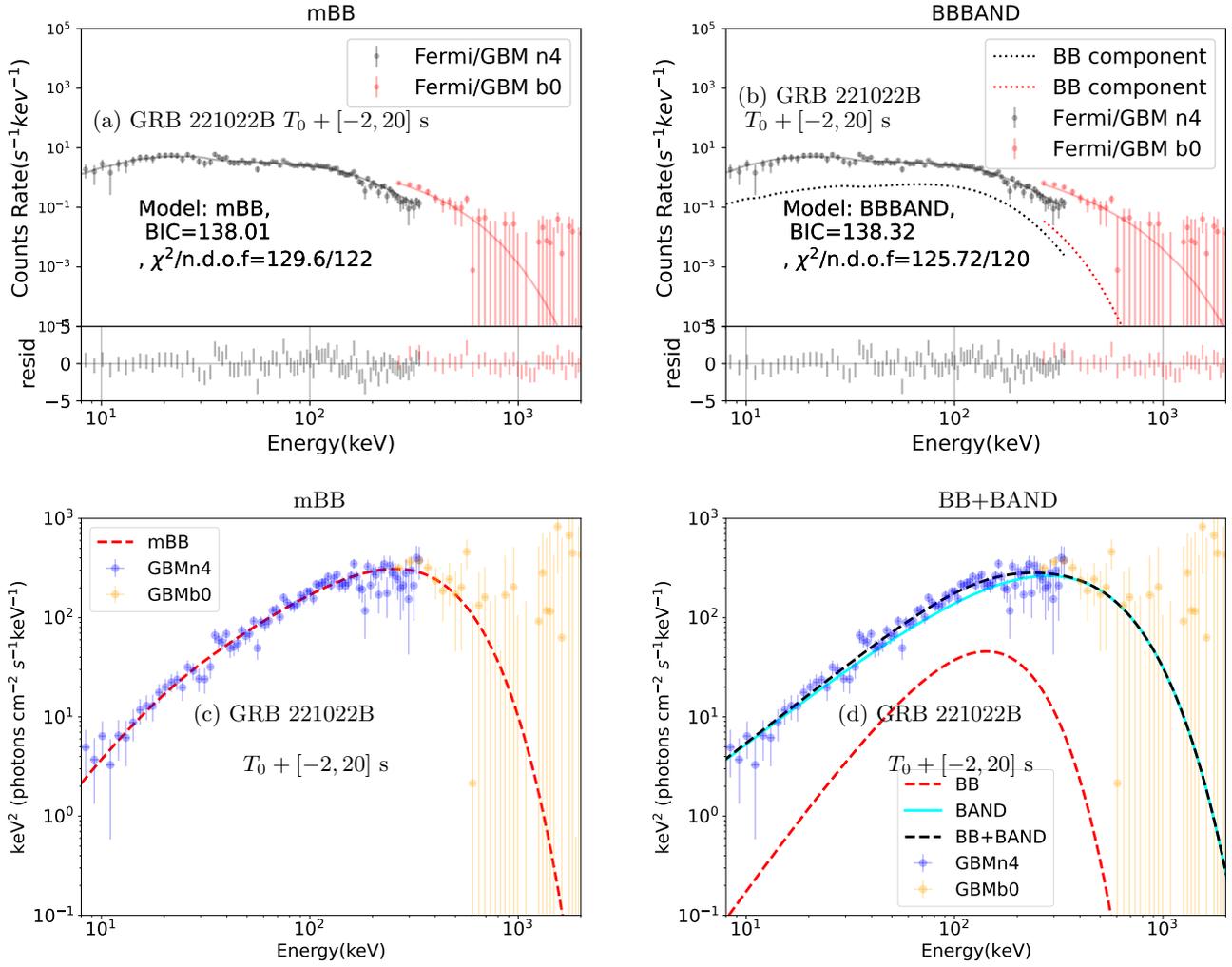
\begin{figure*}
\begin{center}
 \centering
  \includegraphics[width=0.5\textwidth]{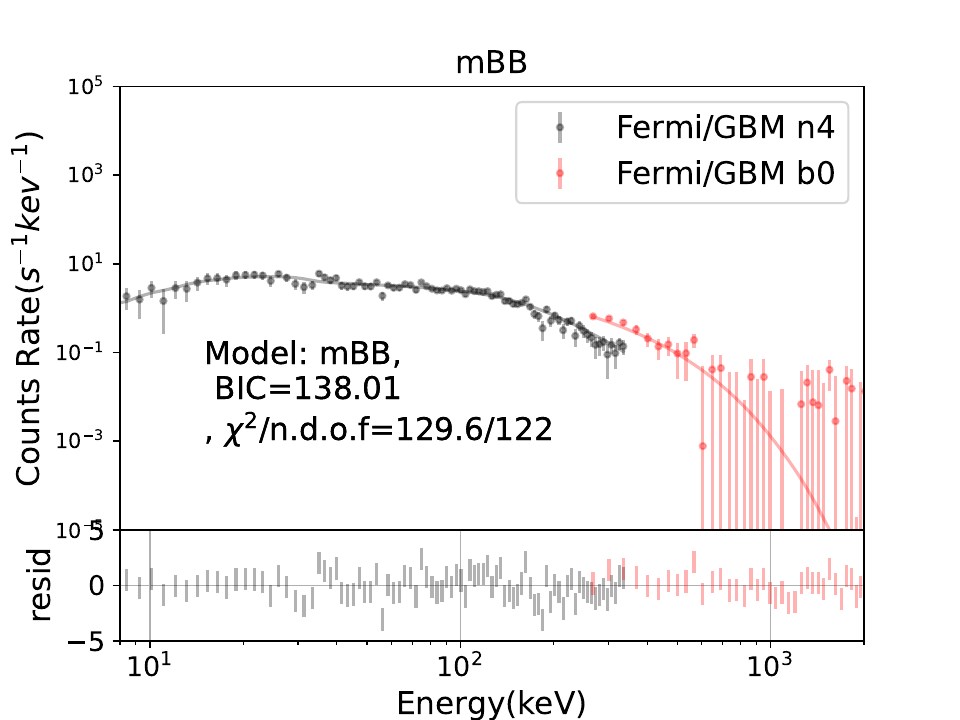}\put(-220,130){(a) GRB 221022B $T_0+[-2, 20]$ s}
  \includegraphics[width=0.5\textwidth]{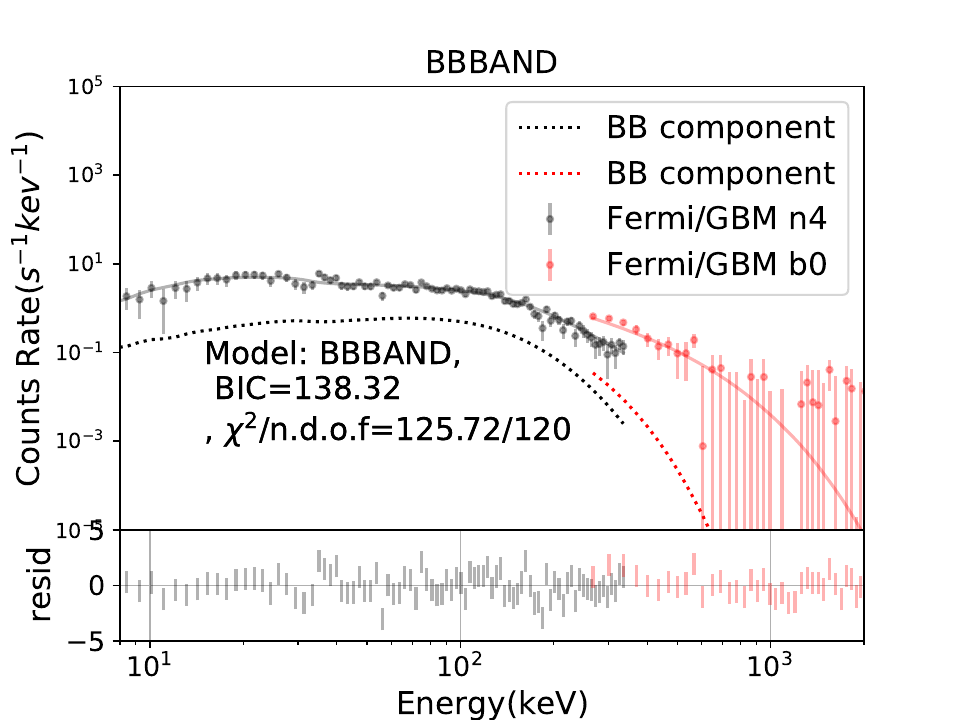}\put(-220,140){(b) GRB 221022B}
  \put(-220,130){ $T_0+[-2, 20]$ s}\\
  \includegraphics[width=0.5\textwidth]{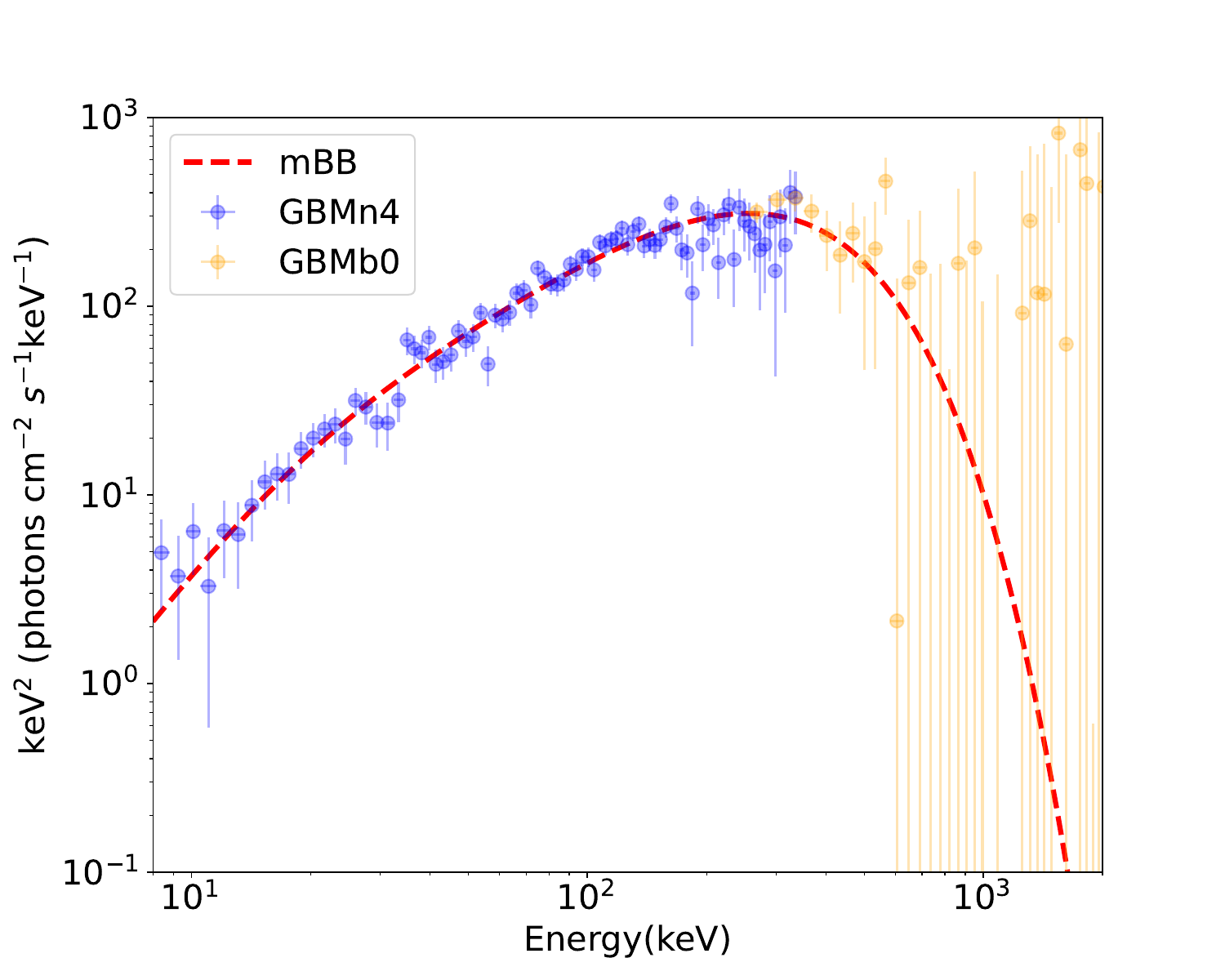}\put(-180,100){(c) GRB 221022B}\put(-140,185){mBB} \put(-160,80){$T_0+[-2, 20]$ s}
  \includegraphics[width=0.5\textwidth]{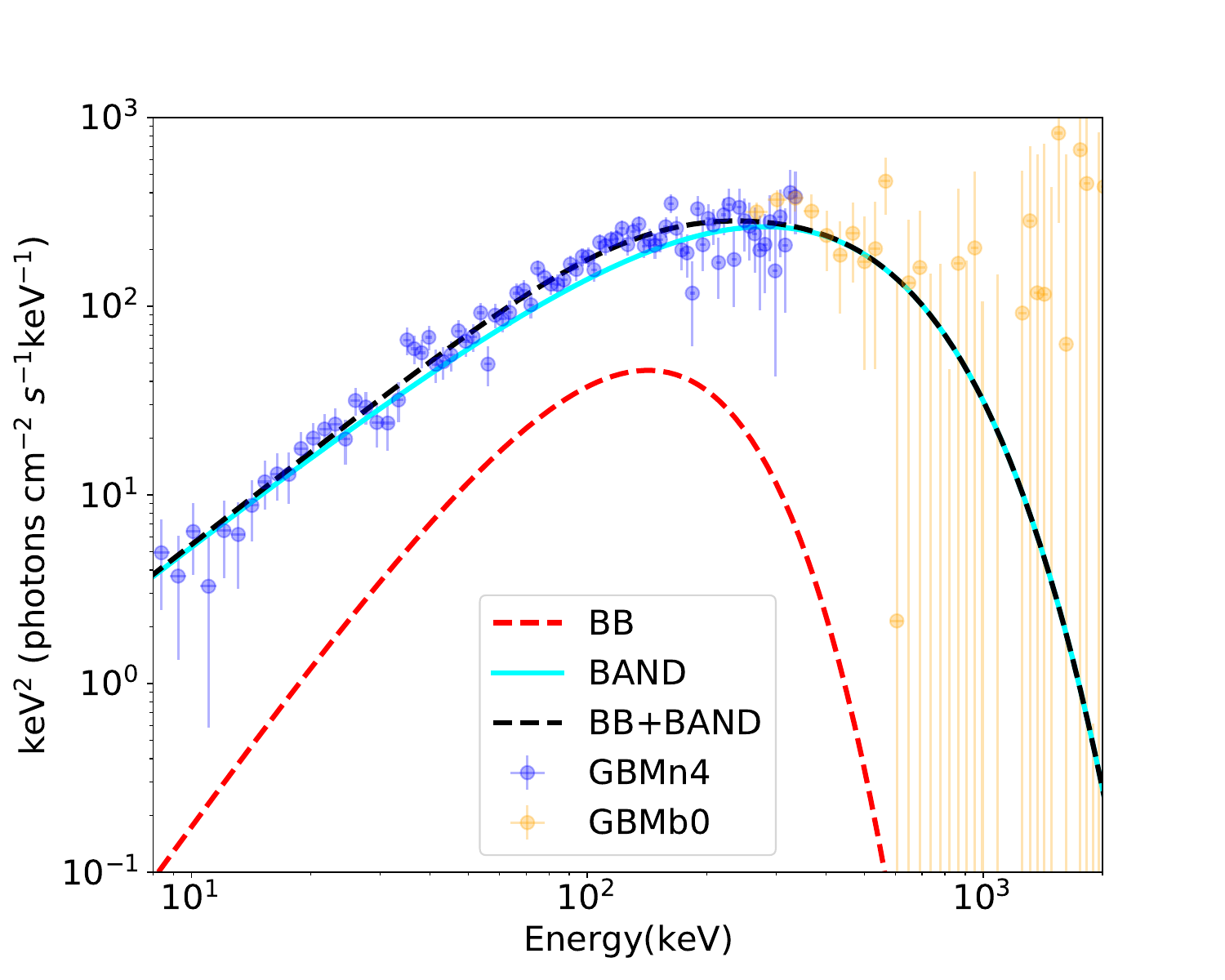}\put(-180,100){(d) GRB 221022B} \put(-160,80){$T_0+[-2, 20]$ s}\put(-140,185){BB+BAND}\\
\caption{\label{fig:GRB221022B_m2_20}(a)-(b) Spectra of $T_0+[-2, 20]$ s in GRB 221022B and fit results of BB+BAND and mBB models, (c) and (d) are model shapes in the $E^2N(E)$ form. }
\end{center}
\end{figure*}

 To confirm that the emission is dominated by quasi-thermal component, the combined model of BB+ BAND is used in the modeling. We find in the first 30 s, $\alpha\sim-0.4$ is obtained for the component described with BAND function, greater than the synchrotron death line. To prove this, we take the spectrum in $T_0+[-2, 20]$ s as an example. Figure~\ref{fig:GRB221022B_m2_20} shows the time-integrated spectrum of $T_0+[-2, 20]$ s and fit results with BB+BAND and mBB. The fit results are listed as below,
 \begin{itemize}
     \item BB+BAND: for BB, $kT=36.3\pm12.0$ keV and flux $F_{\rm BB}=(0.10\pm0.04)\times10^{-6}$ erg cm$^{-2}$ s$^{-1}$; for BAND: $\alpha=-0.35\pm0.05$, $\beta=-9.68\pm0.72$, $E_{\rm p}=280.9\pm10.6$ keV and $F_{\rm BAND}=0.87\pm0.07$ erg cm$^{-2}$ s$^{-1}$. BIC=138.3 with a freedom degree of 120;
     \item mBB: for mBB, $m=0.49^{+0.13}_{-0.23}$, $kT_{\rm min}=5.5^{+3.1}_{-2.4}$ keV, $kT_{\rm max}=105.0^{+14.3}_{-7.9}$ keV, $F_{\rm mBB}=0.94^{+0.06}_{-0.04}\times10^{-6}$ erg cm$^{-2}$ s$^{-1}$;  BIC=138.0 with a freedom degree of 122.
 \end{itemize} 

The modeling with BB+BAND seems also well, however, it is not better than that with mBB by comparing these two BIC values. mBB is preferred with a larger freedom degree. Moreover, $\alpha$ of BAND in BB+BAND $\sim$ -0.4, still well above -2/3. Thus the case is similar to those of two control samples discussed in Section~\ref{sec:test}. $F_{\rm BB}$ in BB+BAND model is very small compared with the total flux. In the following time, the case is similar and the fit quality is not better than those only with empirical models, so we do not show the fit results here.

 An extra component denoted by a PL function is added on the mBB model, to see if there exists non-thermal contribution in the QT-dominated emission. However, for the first and second bins (from $T_0-2$ s to 20 s), and the last two bins (from $T_0+30$ s to 50 s), the fit quality are not better with combined model of mBB +PL, or the fit results are not reasonable (e.g., $m<-1$).  For the former time bins with large $\alpha$, it may be because there is not evident contribution from extra non-thermal (NT) emission; for the latter with much smaller $\alpha<-2/3$, it is dominated by the NT emission. Around the peak flux from $T_0+20$ s to 30 s, we find that the fit results with an extra PL component are better than that only with mBB, especially in $T_0+25$ s to 30 s with $\Delta$BIC=7.9 compared with that only with mBB model. Thus, it seems that in general, evolution from QT emission to NT emission occurs during the burst, as shown in Figure~\ref{fig:LC_GRB221022B} (b).
 
 %\begin{longrotatetable}
\begin{deluxetable*}{lccccccccccc}
\tabletypesize{\tiny}
%\tabletypesize{\small}
\tablewidth{0pt}
\tablecaption{The time-integrated and time-resolved results of GRB 221022B.\label{tab:fitresTR}}
\tablehead{ 
\colhead{Time bins}
&\colhead{Model}
&\colhead{$m$}
&\colhead{$kT_{\rm min}$}
&\colhead{$kT_{\rm max}$} 
&\colhead{$F_T$} 
&\colhead{$\alpha$} 
&\colhead{$\beta$}
&\colhead{$E_{\rm p}$}
&\colhead{$F_{NT}$}
&\colhead{BIC} &\colhead{$\frac{\chi^2}{ndof}$}\\
\colhead{(s)} 
&\colhead{} 
&\colhead{} 
&\colhead{(keV)}
&\colhead{(keV)}
&\colhead{(10$^{-6}$ erg cm$^{-2}$ s$^{-1}$)} &\colhead{}
&\colhead{}
&\colhead{(keV)} 
&\colhead{(10$^{-6}$ erg cm$^{-2}$ s$^{-1}$)} &\colhead{}
&\colhead{}%\\\hline
}
\startdata
[-2,50]&BAND & & & &  &-0.63$^{+0.03}_{-0.03}$ &-3.47$^{+4.03}_{-0.72}$ &274.4$^{+15.0}_{-10.7}$ &1.55$^{+0.03}_{-0.05}$ &251.9 &$\frac{242.3}{242}$\\
&CPL& & & &  &-0.66$^{+0.02}_{-0.02}$ &  &281.8$^{+9.0}_{-9.0}$ &1.45$^{+0.02}_{-0.04}$ &272.3 &$\frac{265.2}{243}$\\
&mBB &0.06$^{+0.03}_{-0.03}$ &4.2$^{+0.5}_{-0.5}$ &131.8$^{+9.3}_{-5.2}$ &1.42$^{+0.03}_{-0.03}$ &  &  &  &  &264.7 &$\frac{255.2}{242}$\\
&mBB+PL &-0.13$^{+0.08}_{-0.08}$ &12.4$^{+5.9}_{-6.5}$ &170.5$^{+140.0}_{-37.6}$ &1.31$^{+0.05}_{-0.10}$ &-1.69$^{+0.07}_{-0.26}$ &  & &0.66$^{+0.16}_{-0.55}$ &327.2 &$\frac{312.9}{240}$\\\hline\hline
[-2,15]&BAND& & & &  &-0.07$^{+0.14}_{-0.12}$ &-6.51$^{+2.00}_{-5.10}$ &234.1$^{+17.5}_{-17.5}$ &0.63$^{+0.04}_{-0.04}$ &74.4 &$\frac{66.0}{122}$\\
&CPL& & & &  &-0.14$^{+0.14}_{-0.11}$ &  &242.2$^{+15.4}_{-20.5}$ &0.63$^{+0.05}_{-0.04}$ &72.4 &$\frac{66.1}{123}$\\
&mBB&0.62$^{+0.18}_{-0.18}$ &5.6$^{+11.4}_{-3.3}$ &98.1$^{+10.0}_{-13.0}$ &0.63$^{+0.04}_{-0.04}$ &  &  &  &  &75.6 &$\frac{67.2}{122}$\\
&mBB+PL&-1.65$^{+0.33}_{-0.35}$ &27.7$^{+2.0}_{-2.0}$ &1840.1$^{+50.6}_{-101.3}$ &0.66$^{+0.05}_{-0.13}$ &-1.78$^{+0.11}_{-0.36}$ &  & &0.23$^{+0.04}_{-0.08}$ &86.9 &$\frac{74.3}{120}$\\\hline
[15,20]&BAND& & & &  &-0.44$^{+0.05}_{-0.08}$ &-6.65$^{+2.40}_{-2.20}$ &259.2$^{+16.0}_{-16.0}$ &2.03$^{+0.08}_{-0.10}$ &116.1 &$\frac{107.7}{122}$\\
&CPL& & & &  &-0.46$^{+0.08}_{-0.06}$ &  &263.2$^{+14.1}_{-17.6}$ &2.02$^{+0.08}_{-0.11}$ &113.8 &$\frac{107.5}{123}$\\
&mBB&0.31$^{+0.08}_{-0.10}$ &3.4$^{+2.0}_{-1.8}$ &112.0$^{+10.4}_{-12.4}$ &1.97$^{+0.09}_{-0.07}$ &  &  &  &  &118.1 &$\frac{109.7}{122}$\\
&mBB+PL&-0.35$^{+1.89}_{-0.12}$ &16.8$^{+18.4}_{-9.8}$ &151.6$^{+140.0}_{-39.9}$ &1.90$^{+0.11}_{-0.08}$ &-1.82$^{+0.19}_{-0.11}$ &  & &0.66$^{+0.09}_{-0.09}$ &128.7 &$\frac{116.1}{120}$\\\hline
[20,25]&BAND& & & &  &-0.54$^{+0.03}_{-0.05}$ &-2.97$^{+0.30}_{-0.10}$ &339.7$^{+26.5}_{-22.1}$ &6.64$^{+0.17}_{-1.25}$ &156.9 &$\frac{148.5}{122}$\\
&CPL& & & &  &-0.57$^{+0.02}_{-0.03}$ &  &357.4$^{+11.3}_{-15.1}$ &5.88$^{+0.16}_{-0.11}$ &164.4 &$\frac{158.1}{123}$\\
&mBB&0.10$^{+0.05}_{-0.05}$ &5.6$^{+0.8}_{-0.8}$ &170.5$^{+10.5}_{-9.3}$ &5.83$^{+0.15}_{-0.15}$ &  &  &  &  &164.3 &$\frac{155.9}{122}$\\
&mBB+PL&-0.11$^{+0.08}_{-0.13}$ &11.9$^{+1.9}_{-1.8}$ &199.5$^{+21.7}_{-15.8}$ &5.64$^{+0.25}_{-0.17}$ &-1.76$^{+0.20}_{-0.08}$ &  & &1.16$^{+0.21}_{-0.21}$ &159.6 &$\frac{147.0}{120}$\\\hline
[25,30]&BAND& & & &  &-0.57$^{+0.05}_{-0.07}$ &-3.39$^{+0.40}_{-0.30}$ &255.8$^{+21.7}_{-18.1}$ &2.86$^{+0.13}_{-0.13}$ &123.0 &$\frac{114.6}{122}$\\
&CPL& & & &  &-0.61$^{+0.05}_{-0.05}$ &  &271.4$^{+15.0}_{-15.0}$ &2.65$^{+0.12}_{-0.09}$ &122.3 &$\frac{116.0}{123}$\\
&mBB&-0.06$^{+0.11}_{-0.13}$ &6.1$^{+1.3}_{-0.8}$ &142.3$^{+11.4}_{-11.4}$  &2.67$^{+0.08}_{-0.14}$ &  &  &  &  &126.1 &$\frac{117.7}{122}$\\
&mBB+PL&-0.60$^{+0.49}_{-0.22}$ &14.7$^{+1.7}_{-1.7}$ &237.5$^{+36.7}_{-50.1}$ &2.61$^{+0.23}_{-0.23}$ &-1.80$^{+0.15}_{-0.41}$ &  & &1.07$^{+0.13}_{-0.27}$ &118.2 &$\frac{105.6}{120}$\\\hline
[30,35]&BAND & & & &  &-0.51$^{+0.26}_{-0.24}$ &-2.25$^{+0.30}_{-2.20}$ &114.5$^{+21.4}_{-28.5}$ &0.94$^{+0.09}_{-0.58}$ &69.6 &$\frac{61.2}{122}$\\
&CPL& & & &  &-0.79$^{+0.11}_{-0.15}$ &  &154.1$^{+19.7}_{-23.6}$ &0.68$^{+0.06}_{-0.06}$ &73.9 &$\frac{67.6}{123}$\\
&mBB&-0.85$^{+0.12}_{-0.10}$ &7.2$^{+0.8}_{-1.1}$ &597.4$^{+332.5}_{-524.1}$  &1.03$^{+0.30}_{-0.21}$ &  &  &  &  &81.9 &$\frac{73.5}{122}$\\
&mBB+PL&-1.21$^{+0.30}_{-0.15}$ &12.3$^{+1.9}_{-1.9}$ &2277.6$^{+1244.0}_{-1028.9}$ &0.65$^{+0.32}_{-0.31}$ &-1.76$^{+0.47}_{-0.02}$ &  & &0.63$^{+0.18}_{-0.18}$ &84.0 &$\frac{71.4}{120}$\\\hline
[35,50]&BAND& & & &  &-0.89$^{+0.10}_{-0.17}$ &-2.46$^{+0.30}_{-4.57}$ &118.3$^{+28.1}_{-20.1}$ &0.56$^{+0.05}_{-0.04}$ &91.3 &$\frac{82.9}{122}$\\
&CPL& & & &  &-1.02$^{+0.08}_{-0.10}$ &  &139.6$^{+18.6}_{-16.3}$ &0.49$^{+0.03}_{-0.04}$ &92.7 &$\frac{86.4}{123}$\\
&mBB&-0.61$^{+0.17}_{-0.14}$ &4.3$^{+0.9}_{-0.9}$ &125.0$^{+12.4}_{-37.1}$ &0.51$^{+0.04}_{-0.05}$ &  &  &  &  &89.7 &$\frac{81.3}{122}$
\\
&mBB+PL&-1.02$^{+0.20}_{-0.10}$ &9.0$^{+1.9}_{-2.0}$ &1841.1$^{+1301.4}_{-1537.5}$ &0.43$^{+0.30}_{-0.23}$ &-1.77$^{+0.13}_{-0.11}$ &  & &0.64$^{+0.24}_{-1.29}$ &122.8 &$\frac{110.2}{120}$\\\hline
\enddata
\end{deluxetable*}
%\end{longrotatetable}

\subsection{Jet properties of GRB 221022B}\label{sec:discussion}
The possible origins for the prompt emission of GRB 221022B could be pure hot fireball or hybrid outflow. Some works~\cite[e.g.,][]{2006A&A...457..763G,2013ApJ...764..157B} have predicted a higher $E_{\rm p}$ varying from 1 MeV up to a maximum value of about 20 MeV depending on magnetization fraction~\citep{2013ApJ...764..157B,2015ApJ...802..134B} in case of sub-photosphere magnetic dissipation, which is much larger than that in GRB 221022B. Thus the assumption of sub-photosphere magnetic dissipation is not considered.

\begin{figure*}
\begin{center}  
  \includegraphics[width=0.4\textwidth]{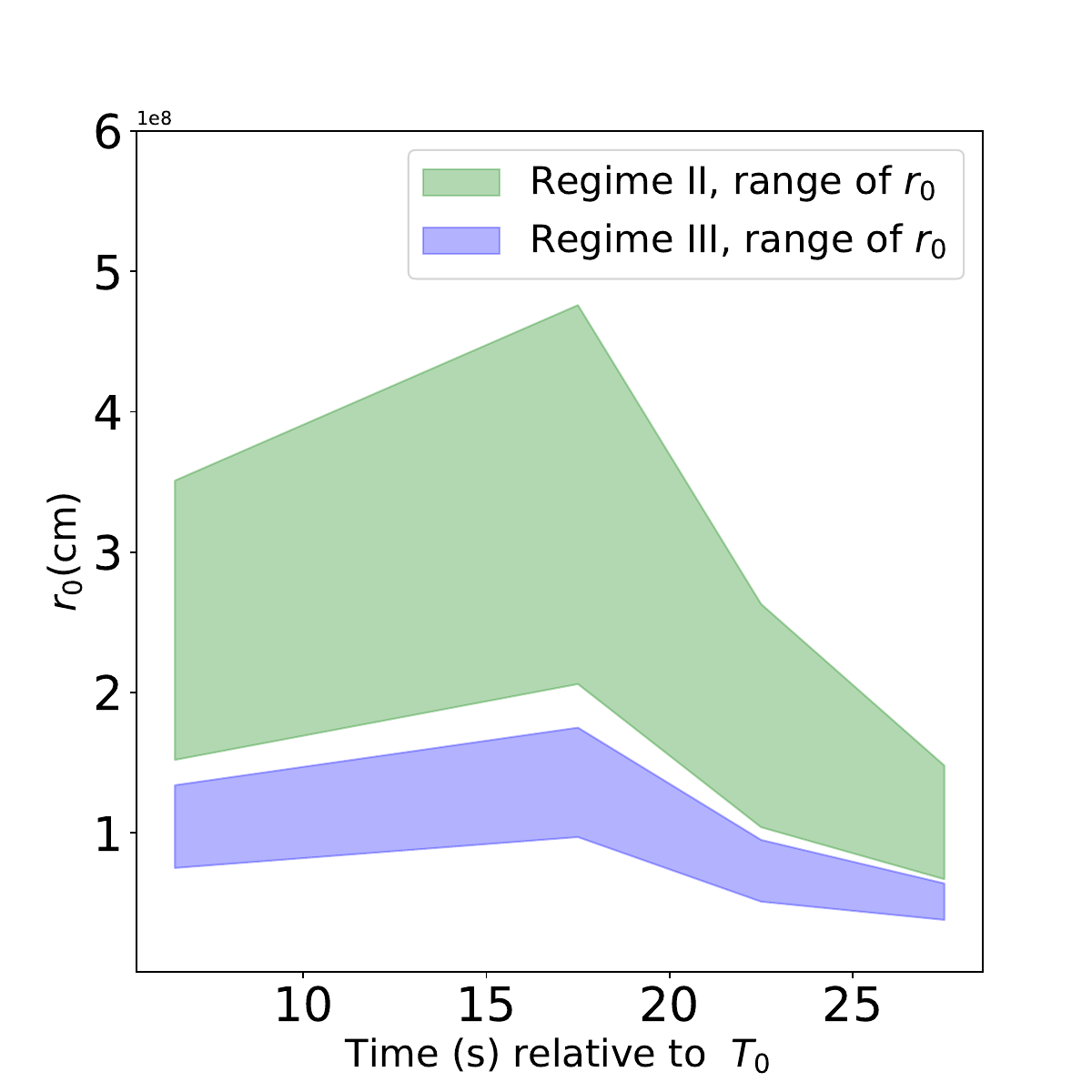}\put(-110,80){(a) GRB 221022B}
  \includegraphics[width=0.4\textwidth]{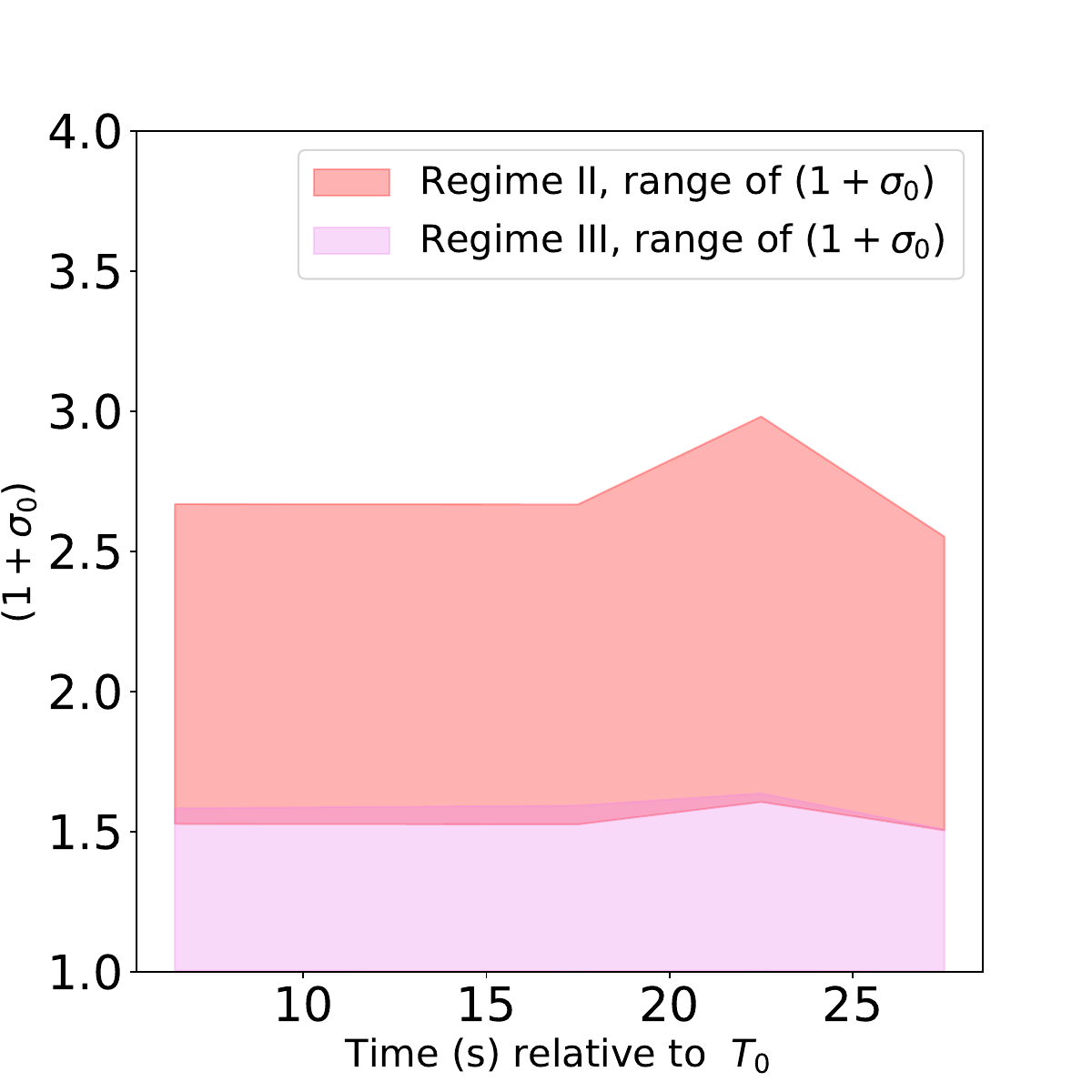}\put(-110,80){(b)GRB 221022B}\\
   \includegraphics[width=0.4\textwidth]{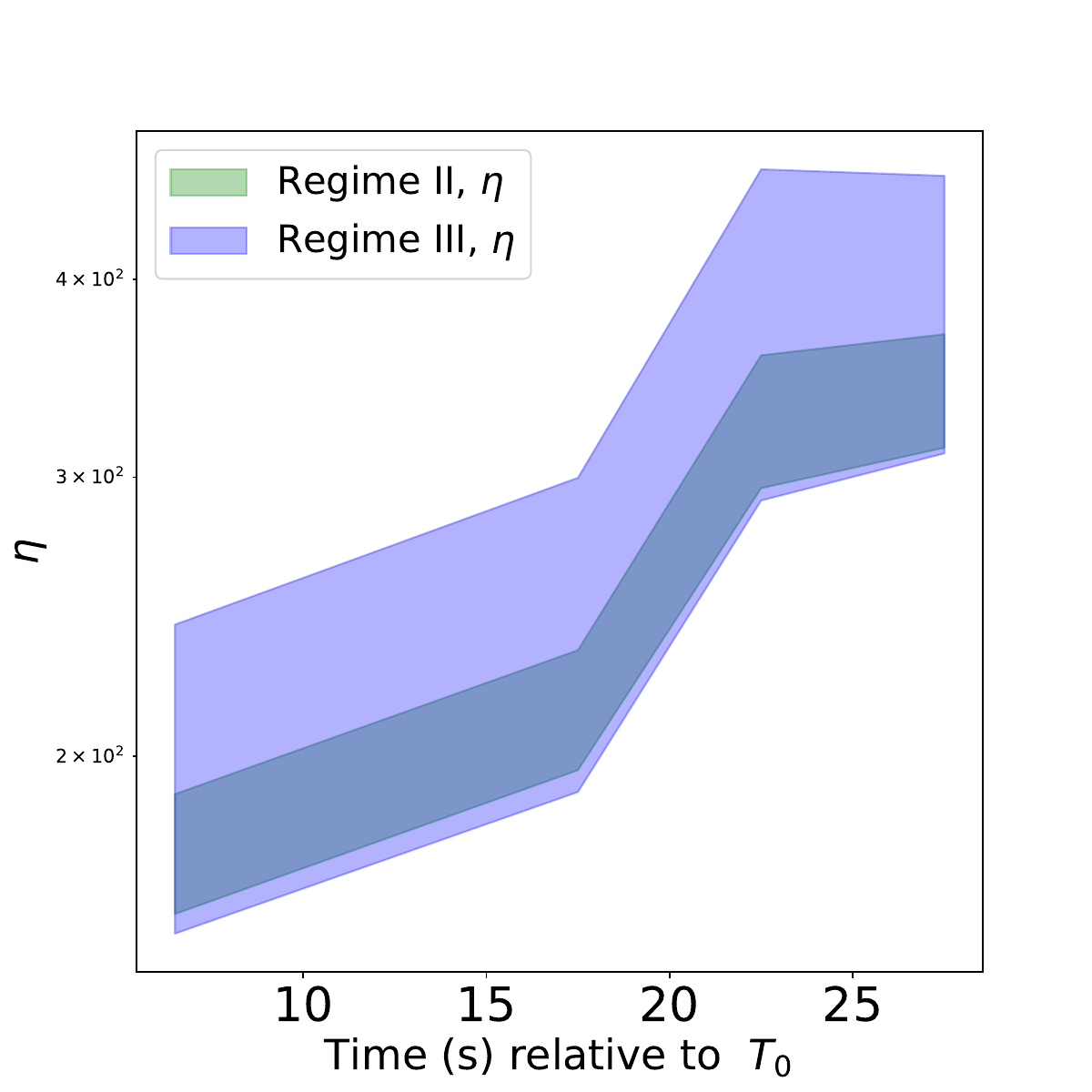}\put(-110,80){(c)GRB 221022B}
  \includegraphics[width=0.4\textwidth]{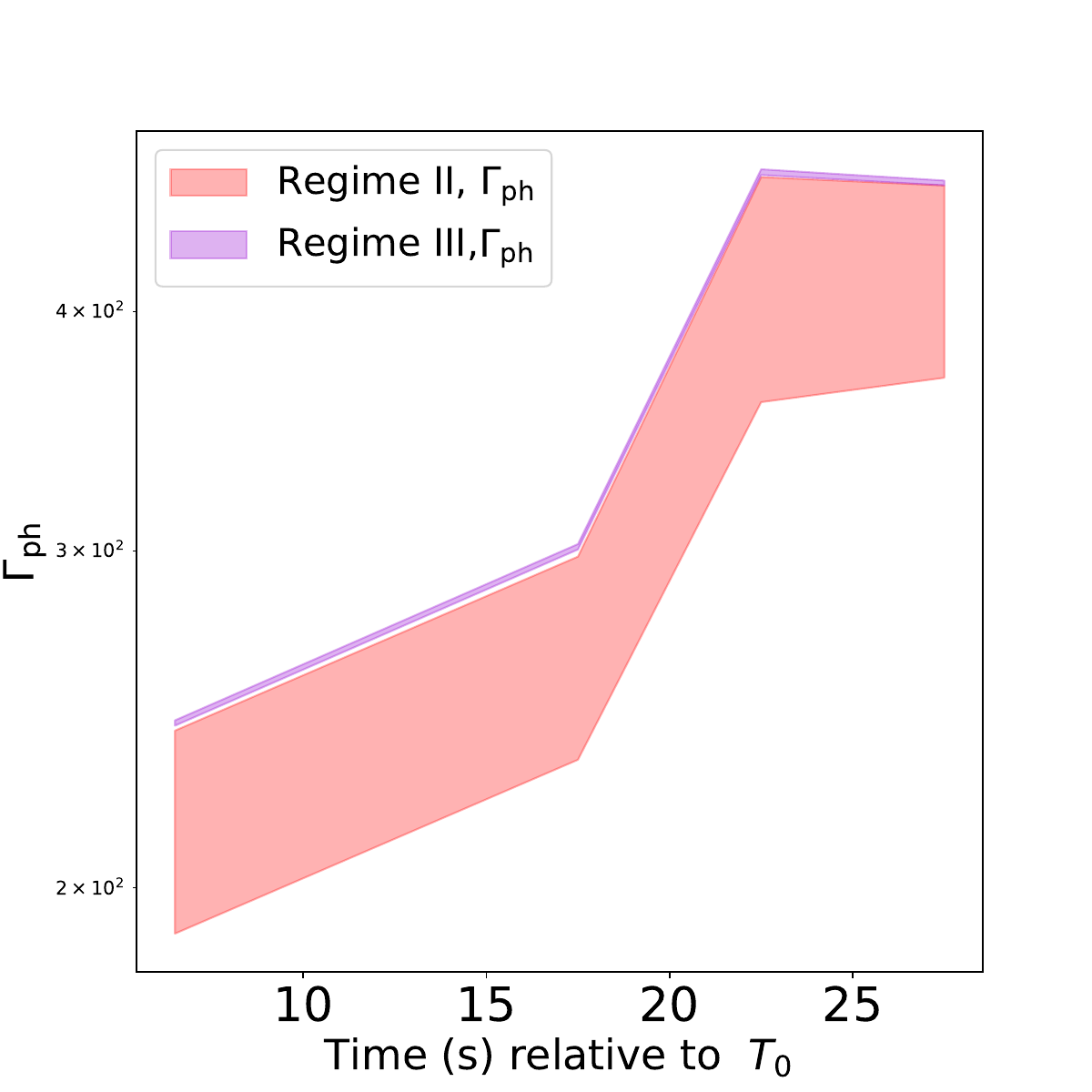}\put(-110,80){(d)GRB 221022B}
\caption{\label{fig:magnetization3}The ranges of $r_0$, $(1+\sigma_0)$, $\eta$, $\Gamma_{\rm ph}$ with $Y=2$ in GRB 221022B.}
\end{center}
\end{figure*}

The diagnosis for magnetization is performed, and the possible regimes are II and III.
Figure~\ref{fig:magnetization3} (a)-(d) shows ranges of $r_0$, $(1+\sigma_0)$, $\eta$, $\Gamma_{\rm ph}$ (the Lorentz factor at $r_{\rm ph}$). In Figure~\ref{fig:magnetization3} (a), blue and green shadows denote the possible values of $r_0$ in the Regimes II and III, where $r_0$ ranges round $10^{8}$ cm, while there exists no reasonable $r_0$ values from $10^{6}-10^{10}$ cm to satisfy Regimes V and VI. The first two bins ($T_0-2$ s to 20 s) could share more or less the same $r_0$ range, and it seems that $r_0$ decreases slightly with time in the following two bins ($T_0+20$ s to 30 s), which seems similar to those of GRB 210121A. $(1+\sigma_0)\simeq1$ could be accepted in Regime III in the $T_0+[-2, 30]$ s, which is similar to the that in GRB 210121A. Moreover, $(1+\sigma_{15})$ is also estimated to be less than 1, corresponding to a coasting regime at $r=10^{14}-10^{15}$ cm, which implies that in the first 30 s, the small NT is likely from IS mechanism, rather than from ICMART mechanism. $\Gamma_{\rm ph}$ and $\eta$ increase with time as shown in Figure~\ref{fig:magnetization3} (c) and (d), which could also support the conclusion that small NT emission is from IS mechanism during the first 30 s. Therefore, we infer that the photospheric emission in GRB 221022B is mainly from the fireball.

$1+\sigma_0$ does not behave an evident increasing trend with time, which implies the possible Poynting flux is not increasing with time, even if it does exist; thus, the small NT emission (as shown in Figure~\ref{fig:LC_GRB221022B} (b) in blue triangles) in the tail of prompt phase ($>T_0+30$ s) is also likely from IS mechanism.

\begin{figure*}
\begin{center}
 \centering
  
  \includegraphics[width=0.45\textwidth]{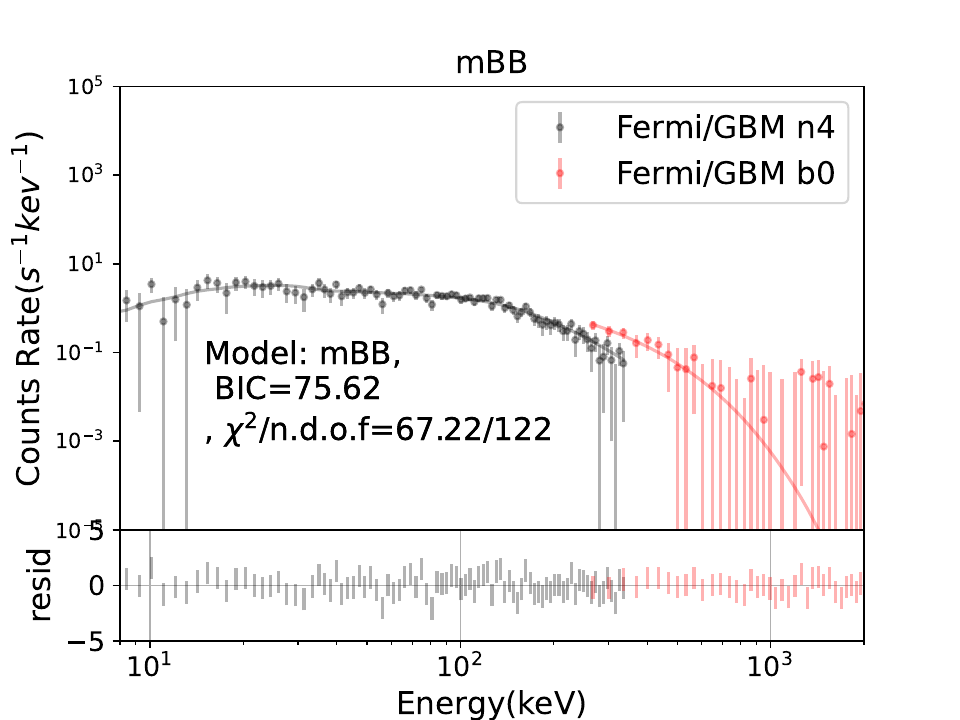}\put(-160,110){(a)GRB 221022B: $T_0$+[-2, 15]s} %mBB bin0
  \includegraphics[width=0.45\textwidth]{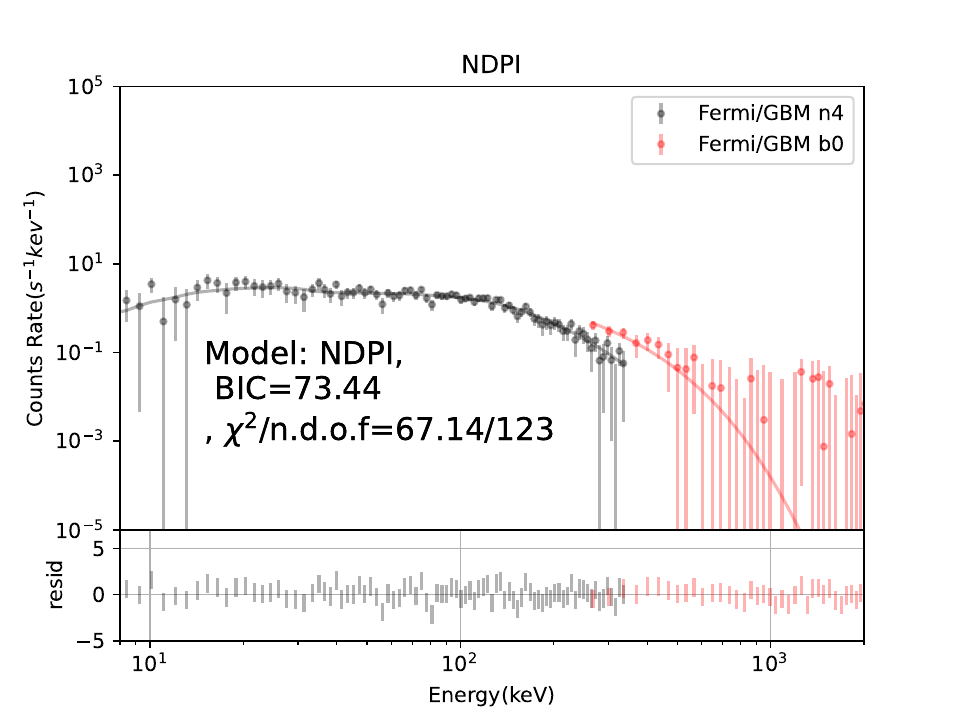}\put(-160,110){(b)GRB 221022B: $T_0$+[-2, 15]s}\\%NDP bin0
  \caption{\label{fig:fitres_bin0} Spectra and fit results in $T_0+[-2, 15]$ s in GRB 221022B with (a) mBB and (b) NDP models.  }
\end{center}
\end{figure*}

Besides, there are other evidences for the origin of fireball. 
% We note that the preferred empirical model for spectra in the beginning of prompt phase, (e.g., $T_0$+[-2, 20] s) is CPL rather than BAND. As proved with the numerical results of non-dissipative hybrid outflow~\citep{2022MNRAS.509.6047M},  the spectrum of prompt emission of non-dissipative magnetic photospheric emission is predicted to be BAND-like with larger $\beta$ than that of CPL function. Thus, we think in the first 20 s, the outflow is dominated by the matter flux, with very small magnetic driven acceleration even if the Poynting flux does exist. 
The mBB+PL model is not preferred in the beginning of the prompt phase, thus, there exists no dissipation from IS mechanism at early time, at least in the first bin, $T_0+[-2, 15]$ s. If the outflow is dominated by hot component without dissipation, the spectra should be described well by probability NDP model~\citep[e.g.,][]{2013A}. To test this, we perform fit with NDP model\footnote{The details of probability NDP model with considering the intermediate photosphere is described in ~\cite{2022ApJ...931..112S}. } on the spectrum of $T_0+[-2, 15]$ s. In this procedure, we use a uniform jet with angle-independent baryon loading\footnote{In the modeling with NDP model in this analysis, we find the extracted values of $p$ and $\theta_{\rm c}$ are large with large uncertainties, where $p$ is the power-law index of the angle-dependent baryon loading and $\theta_{\rm c}$ is the opening angle of the jet. Thus, a uniform jet is assumed, and there are three float parameters ($\eta_0$, $L$ and $r_0$) in the model.}. Therefore, there are only three parameters $r_0$, $\eta$ and $L_{\rm w}$. The fit results is determined to be $\log r_0=8.35^{+0.03}_{-0.03}$ cm, $\eta=227.2^{+100.5}_{-90.3}$ and $\log L_{\rm w}$\footnote{Note this wind luminosity is that in which the probability emission of photons and an opening angle$\sim5/\eta$ are considered.}$=49.67^{+0.02}_{-0.02}$ erg s$^{-1}$ with BIC=73.4. The spectrum is well described by the NDP model as shown in Figure.~\ref{fig:fitres_bin0} (b). The fit quality is even better than that of mBB model with $\Delta$ BIC of 2.2 and a larger freedom degree (BIC is 75.6 and a freedom degree of 122 with mBB model, as shown in Table.~\ref{tab:fitresTR}).

\section{conclusion and Summary}\label{sec:conclusion}

 For the QT-dominated emission of which the spectrum is not hump-like, mBB or mBB+NT may be the preferred model than BB+NT in the modeling to the spectrum, as proved in this analysis\footnote{Note that if we find for a GRB, BB+BAND model has a better BIC or better physical explanation, we will take BB+BAND. This is not an absolute criterion. }. To diagnose the magnetization of the outflow with `top-down' approach, we use a characteristic temperature with the corresponding flux from modeling with mBB or mBB+NT models. 
 For GRB 210610B, the diagnoses based on these two modelings provide similar conclusions qualitatively. However, the estimations based on modeling with mBB (or mBB+NT) may provide more reasonable physical explanations. This implies that the impacts from the GRB jet structure and the geometrical broadening on the observed spectrum should not be ignored in data analysis. Based on this consideration, it is proposed that the jet of GRB 210610B has a moderate magnetization.
 
 From the analysis of the  two control samples (GRB 210610B and 210121A), it seems that a criterion (whether $1+\sigma_0\simeq1$ could be ruled out) could be given to distinguish pure hot fireball from hybrid jet based on the modeling with mBB. However, conservatively, we think it may be difficult to distinguish between these two cases only with these methods, especially in the case of a mild magnetization. Some other information, e.g. the consistence between the data and physical models, the observations from multi-wavelength/messenger, could also provide more evidence. Based on this consideration, for GRB 221022B, the preferred physical model for the early emission is NDP model, offer an additional evidence for its jet properties. From our analysis, it is proposed that the outflow for GRB 221022B is dominated by hot component. Its prompt emission behaves a typical evolution from thermal to NT, in which the NT emission around the peak flux and the tail of the prompt phase is caused by IS mechanism, due to the increasing Lorentz Factor with time.
 %\textbf{Considering most GRBs are synchrotron emission dominated in the prompt phase, these QT-dominated GRBs that are analyzed can not be representative for most GRBs. However, more than one fourth GRBs has a beginning of photospheric emission (usually the first few pulses) with mainly $\alpha$ becoming smaller~\citep[]{2021ApJS..254...35L}. Namely, the first few pulses are similar to those of GRB 210610B and GRB 221022B. }

%\section{}\label{sec:sum}
  
%TC:ignore
\begin{acknowledgements}
 The authors thank supports from the National Program on Key Research and Development Project (2021YFA0718500) and National Natural Science Foundation of China (grant Nos. 12303052).  The authors are very grateful to the public GRB data of Fermi/GBM and HXMT data. We are very grateful for the comments and suggestions of the anonymous referees and Prof. Shuang-Nan Zhang. This work was partially supported by International Partnership Program of Chinese Academy of Sciences (Grant No.113111KYSB20190020).
\end{acknowledgements}
%\clearpage
\bibliography{GRB221022B}{}
\bibliographystyle{aasjournal}
\end{document}